%% file: JLB_Parspl.tex
\documentclass{article}
\pdfoutput=1
\usepackage{ulem}
\usepackage{jcappub}
\usepackage{graphicx}
\usepackage{color}

\def\lbldef#1#2{\expandafter\gdef\csname #1\endcsname {#2}}

\def\href#1#2{#2}

\usepackage{pifont}
\usepackage{multirow}

\newcommand{\La}{\Lambda}

\newcommand{\beq}{\begin{equation}}
\newcommand{\eeq}{\end{equation}}

\title{
Parameter splitting in dark energy: is dark energy the same in the background and in the cosmic structures?}
\author[a,b]{Jos{\'e} Luis Bernal,}
\author[c,a,d,e,f]{Licia Verde,}
\author[a]{and Antonio J. Cuesta}

\affiliation[a]{Institut de Ci{\`e}ncies del Cosmos (ICCUB), Universitat de Barcelona (IEEC-UB), Mart{\'\i} i Franqu{\`e}s 1, E08028 Barcelona, Spain}
\affiliation[b]{Dept. d'Astronomia i Meteorologia,
Universitat de Barcelona, Mart{\'\i} i Franqu{\`e}s 1, E08028 Barcelona, Spain}
\affiliation[c]{ICREA (Instituci\'o Catalana de Recerca i Estudis Avan\c{c}ats)}
\affiliation[d]{Institute of Theoretical Astrophysics, University of Oslo, 0315 Oslo, Norway}
\affiliation[e]{Radcliffe Institute for Advanced Study, Harvard University,   MA 02138, USA}
\affiliation[f]{ITC, Harvard Smithsonian center for Astrophysics,  MA 02138, USA}

\emailAdd{joseluis.bernal@icc.ub.edu}
\emailAdd{liciaverde@icc.ub.edu}
\emailAdd{ajcuesta@icc.ub.edu}

\abstract{ We perform an empirical consistency test of General Relativity/dark energy by disentangling expansion history and growth of structure constraints.
We replace  each late-universe parameter that describes the behavior of dark energy with two meta-parameters: one describing geometrical information in cosmological probes, and the other controlling the growth of structure. If the underlying model  (a standard $w$CDM cosmology with General Relativity) is correct, that is under the null hypothesis, the two meta-parameters coincide.  If they do not, it could indicate a failure of the model or systematics in the data. We present a global analysis  using  state-of-the-art cosmological data sets which points in the direction that cosmic structures prefer a weaker growth than that inferred by background probes. This result could signify  inconsistencies of the model, the necessity of extensions to it or  the presence of systematic errors in the data.  We examine all these possibilities. The fact that the result is mostly driven by a specific sub-set of  galaxy clusters abundance data, points to the need of a better understanding of this probe.
}

\begin{document}

\maketitle
\hypersetup{pageanchor=true}
\input{sec1}
\input{sec2}
\input{sec3}
\input{sec4}
\input{sec5}

\input{sec6}

\input{sec7}

\acknowledgments
\input{Acknowledgements}

\bibliography{Refs}
\newpage
\appendix
\input{Appendix}

\bibliographystyle{utcaps}
\end{document}

%% file: sec1.tex
\section{Introduction}
	Since the detection of the accelerated  expansion of the Universe in the late 1990s \citep{Riess98,Perlmutter99} a spatially flat Cosmology, with matter content dominated by cold dark matter but with total matter-energy density dominated by a cosmological constant component ($\La$CDM), has been adopted as
 the standard cosmological model. The $\La$CDM model has shown an impressive agreement with the observational data gathered so far.  Over the past 15 years, an unprecedented observational effort, culminating with the latest observations of the Cosmic Microwave Background (CMB) by the Planck satellite \citep{Planck15} and of large scale structure clustering by several surveys \citep{Anderson14, BlakeWiggle}, has tested this model. The model has survived the avalanche of data and its parameters are  exquisitely well constrained, some with below percent precision. The latest data impose stronger than ever constraints on new physics beyond the $\La$CDM model e.g., \cite{Planck15_MG,Planck15_inf, Bellini15, Samushia14} and Refs. therein.
 
 Nevertheless, the cosmological constant  as a non-varying vacuum energy  is highly fine tuned in the absence of a fundamental symmetry that sets the value of this constant to its small observed value e.g.~\cite{Weinberg89}.  Alternative dynamical models or  modifications of the gravitational sector of the theory, have been proposed (see e.g. \cite{Albrecht06, Mortonson14}, and Refs. therein).
 However, the origin of the cosmic acceleration is still unknown. 

In this work we aim to shed some light on the nature of the cosmic acceleration, trying to distinguish  whether it has its origin in dynamical dark energy or modified gravity, using a phenomenological approach.  As first emphasized by \citep{Starobinsky98}, in General Relativity (GR) within minimally coupled dark energy models, the expansion history fully determines the growth history.  Conversely, if cosmic acceleration is to be explained with modifications of the gravitational sector, the growth history inferred from the expansion history using General Relativity would not necessarily fit the observed growth of cosmic structures.  In  the pioneering work of Ref. \citep{Ishak06} it was proposed using this fact to perform a consistency test of GR/dark energy.
 
Therefore, we use an approach that we refer to as {\it parameter splitting}.
  This technique is based on the replacement of a cosmological parameter with two independent meta-parameters, --hence the name splitting--. 
 In our work,  one of the two effective parameters is used to fit cosmological observations that probe the  expansion history of the Universe  while the other one  is determined by  the growth of structures. Ideally, in a scenario where only statistical errors are present, if GR holds at cosmological scales and the acceleration of the expansion of the Universe is caused by  minimally coupled dark energy, the two (split) parameters must coincide. In this sense, parameter splitting offers a null test. On the contrary, if modifications of GR cause the cosmic acceleration, the growth history is not uniquely determined from the expansion history and the two split parameters need not to be identical.

Parameter splitting is a general but powerful technique to check the consistency of a model and, in this case, to shed some light on the nature of dark energy. Its advantage is that, by offering a null test and resorting to meta-parameters, it is model-independent, or at least less sensitive to the choice of an alternative model than traditional approaches.
If the null test is not failed, there is no reason to claim that  new physics is needed. On the other hand, if the null test is failed, in the absence of systematic errors, this indicates a mismatch between the growth history derived from expansion history using GR and the observed growth of structures, offering insights into the new physics needed by the standard  $\La$CDM and the nature of dark energy.  
However, the presence of  unaccounted systematic errors can also lead to a failure of the split parameters null test; hence parameter splitting is also sensitive to systematic errors. Nonetheless, by performing the analysis with  different, suitably chosen, combinations of data sets, it allows us  to disentangle systematics from new physics.

Parameter splitting has been used in cosmology in different contexts in Refs. \citep{Zhang05, Chu05, Abate08}, but it was applied for the first time to the dark energy parameters in \citep{Wang07}, where the dark energy density parameter $\Omega_{\rm DE}$ and the dark energy equation of state parameter $w$ (which is $w=-1$ in the $\La$CDM model) is split into $\Omega_{\rm DE}^{\rm geom}$ and $\Omega_{\rm DE}^{\rm growth}$ and $w_{\rm geom}$ and $w_{\rm growth}$ (both considered constants), respectively. No evidence was found of any discrepancy between these split parameters. Since this work appeared, there has been a huge increase both in precision and quantity of data, so an update of Wang's work (Ref.~\citep{Wang07}) is necessary. Indeed, this technique has recently been used in the dark energy parameters in \citep{Ruiz15}. When the authors of Ref. \cite{Ruiz15} perform the split in two parameters simultaneously, $w$ and $\Omega_{\rm DE}$, they find a tension of 3.3$\sigma$ between $w_{\rm geom}$ and $w_{\rm growth}$.  This result was found to be mostly driven by Redshift Space Distortions (RSD) data. They also find that the  tension disappears when they allow the  sum of neutrino masses ($\sum m_\nu$) to vary, but the resulting $\sum m_\nu$ is too high compared to the limits set by the latest observations \citep{Palanque15}.

Here we  use parameter splitting technique and  build on  Refs.~\cite{Wang07,Ruiz15}  to  test the internal consistency of the standard cosmological model and identify any specific data set which shows tension with the rest.
 We use the latest cosmological data in full (i.e., compared to \cite{Ruiz15} we use the full CMB likelihood rather than the compressed constraints on the CMB peak locations, the temperature and polarization power spectrum data from Planck 2015 rather than the temperature only power spectrum from Planck 2013, and the latest Supernovae data) and  a different methodology (i.e., without using priors from previous data sets or top hat limits to the fundamental parameters). Finally, we perform the test of adding neutrino physics beyond the standard model to ascertain if adding these extra parameters (effectively representing new physics beyond $\La$CDM) is favored by the data.
 
In Sec. \ref{sec:Data} we list the data we use in our analysis. In Sec. \ref{sec:method} we present our methodology and the parameters we consider. In Sec. \ref{sec:Results} we present our results to discuss them in Sec. \ref{sec:NewPhys} and \ref {sec:discussion}. Finally in Sec.\ref{sec:conclusions} the conclusions are summarized.

%% file: sec2.tex
\section{Data}\label{sec:Data}
In this section we discuss the observational data we use in our analysis. We separate the cosmological probes depending on whether they are sensitive to the expansion history of the Universe or to structure growth (table \ref{tab:DATA}).  The power of the meta-parameters approach is that we can use in the same analysis  observables   that probe growth  and observables that probe the  expansion history. For example, as detailed in Sec. \ref{sec:method},  the first set of probes  will be sensitive to the growth meta-parameters (and not the geometry ones) while the second will be sensitive to the geometry meta-parameters (and not the growth ones).

We do not include in this work the  weak lensing power spectrum measurements, such as for example the widely used ones from CFHTLens  survey \citep{Kilbinger13}. In weak gravitational lensing by galaxies, the signal coming from geometry and that coming from growth are intertwined  and not easy to separate out. We leave this to forthcoming work.

\begin{table}
\small
\begin{center}
\begin{tabular}{cccc}
\hline
Cosmological Probe	& Measurement	& Geometry	& Growth \\
\hline
CMB		& high $\ell$ in TT, TE and EE power spectra	& \checkmark		& \\
CMB		& low $\ell$ in TT, TE and EE power spectra	& 			& \checkmark \\
CMB		& $40\leq \ell \leq 400$ lensing power spectrum	& 			& \checkmark \\
SNeIa	& $D_L$						& \checkmark		& \\
BAO		& $D_V/r_{\rm s}$, $D_A/r_{\rm s}$ and $c/(Hr_{\rm s})$	& \checkmark		& \\
Clusters		& $\Omega_{\rm M}^\beta \sigma_8$	&		& \checkmark \\
RSD		& $f\sigma_8$				& 			& \checkmark \\
\hline
\end{tabular}
\end{center}
\caption{\footnotesize
Summary of the cosmological probes that we use specifying the measurements included in the analysis and specifying if they are used to constrain either geometry or growth. For CMB ``low $\ell$'' means $\ell < 30$ and ``high $\ell$'' means $30 \leq \ell < 2508$ for TT data and $30 \leq \ell \leq 1996$ for TE and EE data.
}
\label{tab:DATA}
\end{table}

\subsection{Cosmic Microwave Background}\label{sec:CMB}
We use the Planck 2015 temperature (TT), polarization (EE) and the cross correlation of temperature and polarization (TE) \citep{Planck15_likelihood} together with the CMB lensing power spectrum \citep{Planck15_Param}. The CMB acoustic peaks provide an excellent standard ruler and set strong constraints on the composition of the Universe. We choose to use the whole shape of the CMB angular power spectrum in order to perform a more complete analysis without the necessity of using strong priors. The high multipoles ($\ell\geq 30$) of the spectra are used here to constrain the expansion history. The low multipoles  ($2\leq \ell \leq 29$), which are dominated by Integrated Sachs-Wolfe effect, are used to constrain growth. Gravitational lensing effects in the CMB spectra are sensitive to large scale structures at intermediate redshift, hence we use the lensing likelihood, in the multipole range $40\leq \ell \leq 400$ of the TT, TE and EE power spectra, to constrain growth parameters too. In order to explore the dependence of the results on Planck 2015 data, we also perform the analysis using Planck 2013 data \citep{Planck13_Param} instead of Planck 2015 data. The main difference between Planck 2013 and Planck 2015 is that in the former the low multipoles polarization data ($\ell \leq 30$) are taken from WMAP9 \citep{Bennett13}. When we use Planck 2013 data we use the high multipoles ($50 \leq \ell \leq 2500$) of the TT power spectrum to constrain geometry and the low multipoles ($\ell\leq 50$) of TT power spectrum, the lensing reconstruction, and WMAP9's low multipoles of EE power spectrum  to constrain growth. Hereinafter, whenever `Planck 2015' or `Planck 2013' is named without further specification, we will refer to the full TT, TE, EE and CMB lensing power spectra, used as explained above, but in the likelihood function -- see sec.3 for more details-- each  is compared to the theory  prediction obtained using the corresponding --growth or expansion-- meta-parameters.

\subsection{Supernovae Type Ia}
Type Ia Supernovae (SNeIa) are one of the best probes to measure the redshift-distance relation, as each supernova (assuming they are standard candles) provides a direct independent measurement of the luminosity distance. Therefore, here we use SNeIa to constrain the expansion history. The dataset used here is the SDSS-II/SNLS3 Joint Light-curve Analysis (JLA) data compilation \citep{Betoule14}. This catalog contains 740 spectroscopically confirmed SNeIa obtained from low redshift samples ($z<0.1$), all three seasons of the Sky Digital Sky Survey II (SDSS-II) ($0.05<z<0.4$) and the three years of the SuperNovae Legacy Survey (SNLS) ($0.2<z<1$) together with a few additional SNeIa at high redshift from HST ($0.8<z<1.2$).

\subsection{Baryon Acoustic Oscillations}
The sound horizon at radiation drag ($r_{\rm s}$) imprints a characteristic scale not only in the acoustic peaks of the CMB power spectrum but also in the large scale structure. The physics which dominates the features in the CMB anisotropies distribution is well understood, so the sound horizon becomes a standard ruler throughout the Universe ($r_{\rm s} \sim 150$ Mpc). That is why Baryon Acoustic Oscillations (BAO) have become one of the most robust and strongest probes of the expansion history of the Universe. We use the following BAO data to constrain the expansion history: the measurement from the Six Degree Field Galaxy Survey (6dF) \citep{Beutler11}, the Main Galaxy Sample of Data Release 7 of Sloan Digital Sky Survey (SDSS-MGS) \citep{Ross15}, the LOWZ and CMASS galaxy samples of the Baryon Oscillation Spectroscopic Survey (BOSS-LOWZ and BOSS-CMASS, respectively) \citep{Anderson14}, and the distribution of the Lyman-$\alpha$ Forest in BOSS (BOSS-Ly$\alpha$) \citep{FontRibera14}. These measurements, and their corresponding effective redshift $z_{\rm eff}$, are summarized in table \ref{tab:BAO}.

\begin{table}
\small
\begin{center}
\begin{tabular}{cccc}
\hline
Survey	& $z_{\rm eff}$	& Parameter	& Measurement\\
\hline
6dF \citep{Beutler11}	& 0.106	& $r_{\rm s}/D_V$	& $0.327\pm 0.015$ \\
SDSS-MGS \citep{Ross15}        & 0.15  & $D_V/r_{\rm s}$       & $4.47\pm 0.16$\\
BOSS-LOWZ \citep{Anderson14}	& 0.32	& $D_V/r_{\rm s}$	& $8.47\pm 0.17$\\
BOSS-CMASS \citep{Anderson14}	& 0.57	& $D_V/r_{\rm s}$	& $13.77\pm 0.13$\\
BOSS-Ly$\alpha$ \citep{FontRibera14}	& 2.36	& $c/(Hr_{\rm s})$		& $9.0\pm 0.3$\\
BOSS-Ly$\alpha$ \citep{FontRibera14}	& 2.36	& $D_A/r_{\rm s}$	& $10.8\pm 0.4$\\
\hline
\end{tabular}
\end{center}
\caption{\footnotesize
BAO data measurements included in our analysis specifying the survey that obtained each measurement and the corresponding effective redshift $z_{\rm eff}$. 
}
\label{tab:BAO}
\end{table}

\subsection{Galaxy clusters}
The abundance of galaxy clusters is a powerful probe of the growth of cosmic structures. However, this probe relies on the calibration of the mass-observable relation, which represents  the biggest  uncertainty.
To fully exploit the strength of galaxy clusters counts as a cosmological probe it is necessary to consider the number of halos within a redshift and mass bin. This computation involves a geometrical determination of the cosmological volume element and is hard and expensive to perform. On the other hand, 
 the cosmological information enclosed in the cluster abundance  is efficiently compressed in a constraint on the combination,
\beq
\sigma_8\left(\frac{\Omega_{\rm M}}{\alpha}\right)^\beta \,,
\label{Omega_sigma8}
\eeq
where $\sigma_8$ is the linear amplitude of fluctuations on scales of $8h^{-1}$ Mpc, $\Omega_{\rm M}$ denotes the matter density parameter and $\alpha$ is the fiducial value for this parameter adopted in each analysis.

In \citep{Ruiz15} the full expression of the cluster abundance is used to constrain both geometry and growth. However, they find that this cosmological probe practically does not constrain the geometry  meta-parameters  but does constrain the growth meta-parameters, especially $\Omega_{\rm DE}^{\rm growth}$ (Figures 11 and 12 of that Ref.). This  demonstrates that cluster abundance carries much less information about geometry than about growth of structures for the adopted choice of model's parameters and meta-parameters. This  consideration justifies our choice of using the parameter combination expressed in eq.\eqref{Omega_sigma8} to constrain  growth meta-parameters and not geometry meta-parameters. With such consideration, we lighten up the computational load of our analysis considerably.

\begin{figure}[h]
\minipage{0.5\textwidth}
\begin{center}
\includegraphics[width=0.9\textwidth]{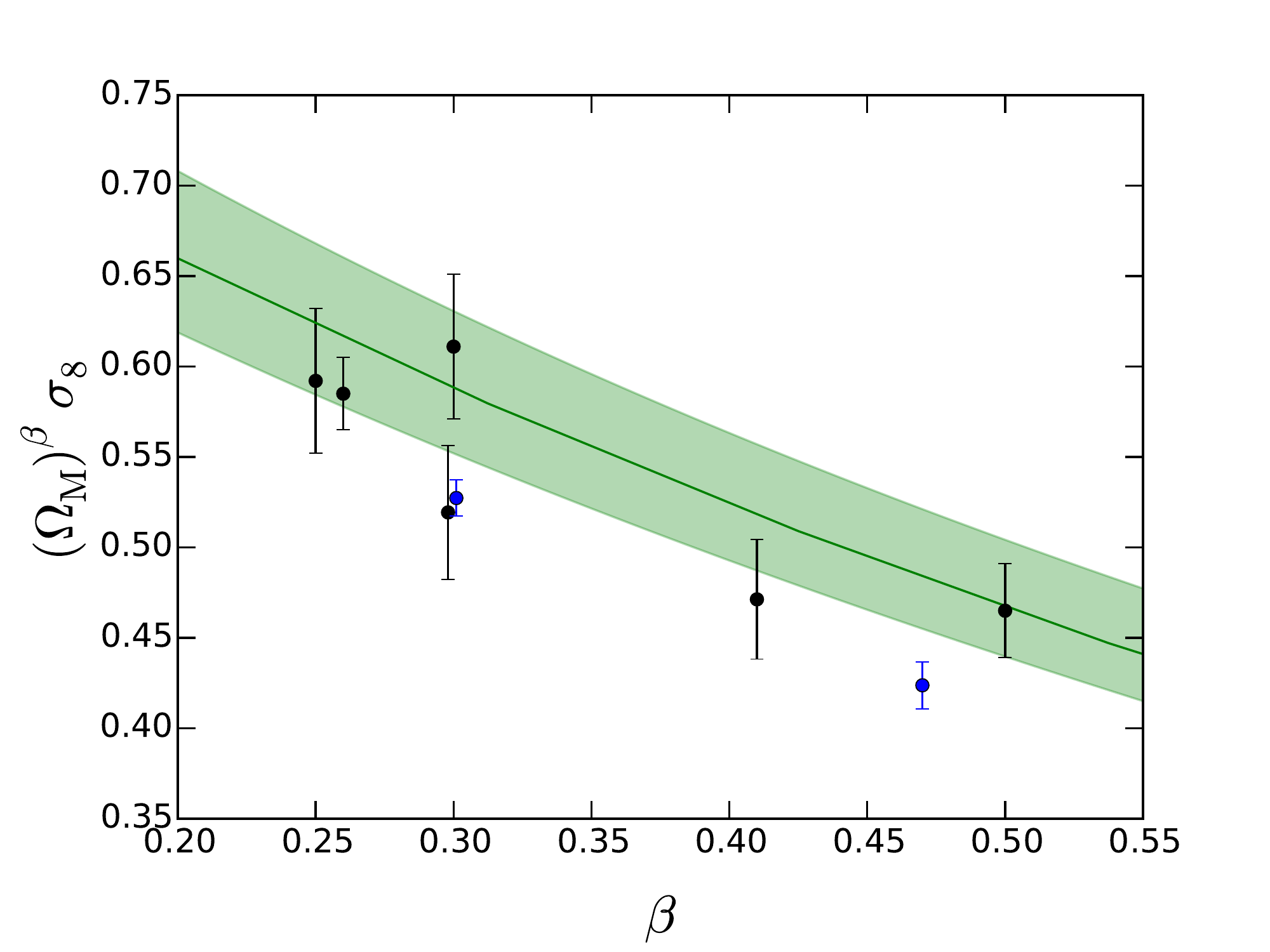}
\end{center}
\endminipage
\minipage{0.5\textwidth}
\begin{center}
\includegraphics[width=0.9\textwidth]{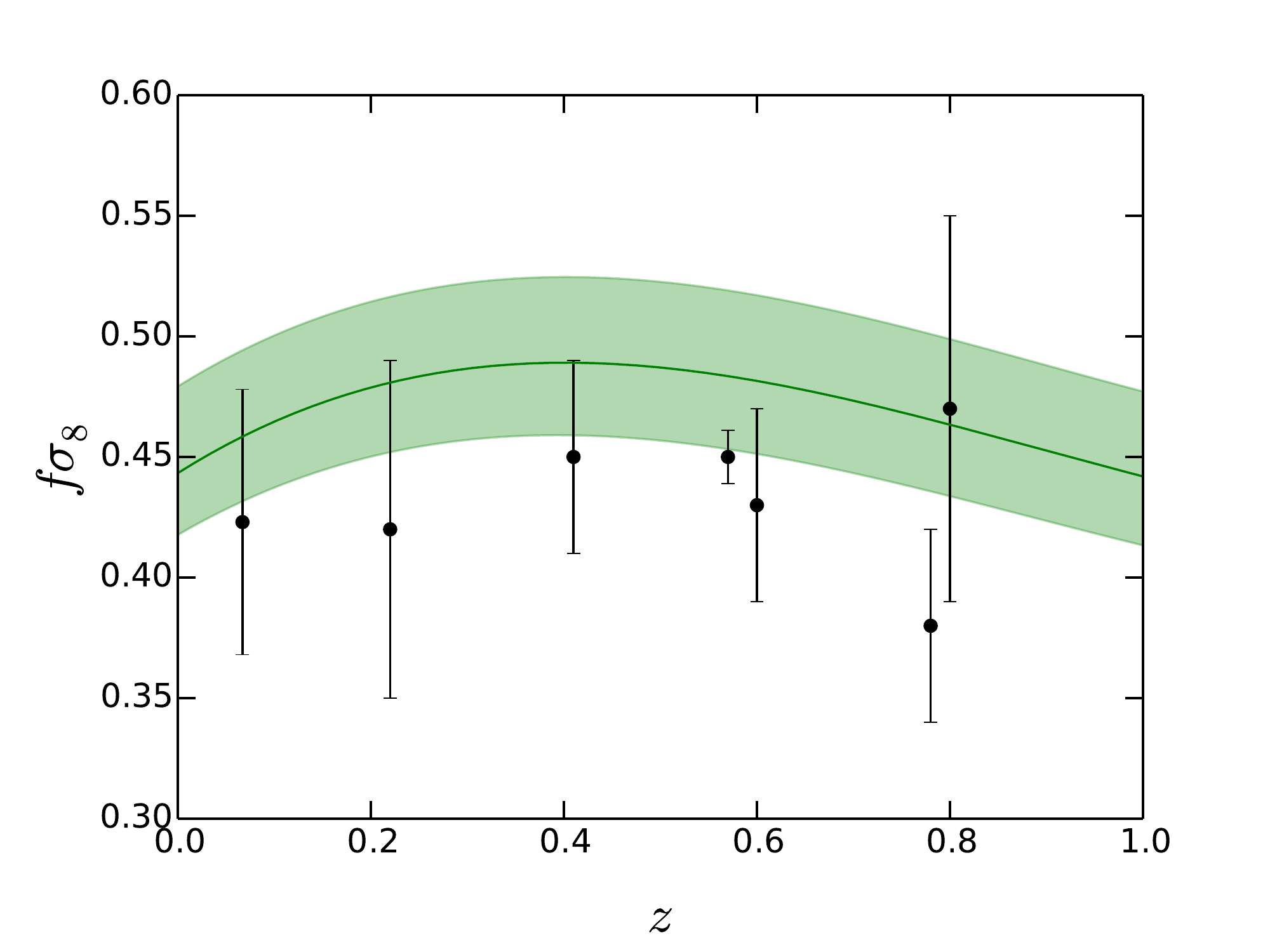}
\end{center}
\endminipage\hfill
\renewcommand{\baselinestretch}{1}
\caption{\footnotesize
\textit{Left}: Parameter combination $\Omega_{\rm M}^\beta\sigma_8$ for the `clusters data set' (blue dots) and the `alternative clusters data set' (black dots) as a function of $\beta$. These values are presented in table \ref{tab:clusters}. \textit{Right}: the RSD data (black dots), shown in table \ref{tab:RSD}, as a function of redshift. In both figures, the shaded regions show the region encompassing the 68\% of the points, as computed from the Planck 2015 public chains \citep{Planck15_Param} (\textit{Planck}TTTEEE+low$\ell$), and the green line is the best fit assuming a $\La$CDM model in that case.
}
\label{fig:Preds_data}
\end{figure}

We use the measurements of table \ref{tab:clusters} to constrain growth.  These are also reported in  the left panel of  Figure \ref{fig:Preds_data}  along with the  $\La$CDM prediction from Planck 2015.
We note that two of the measurements in this list are not consistent with $\La$CDM. These are the ones obtained from  X-ray observations of galaxy clusters made by Chandra \citep{Vikhlinin09} and from the  Sunyaev-Zeldovich effect measured by Planck \citep{Planck13_SZ}. They happen to have the smallest error-bars and thus they dominate in a joint analysis. Therefore we divide the cluster sample into two datasets: one with only the two measurements showing tension with $\La$CDM (which we refer to as `clusters data set' and unless otherwise stated it is the one included in the analysis), and the other including the remaining six measurements (which we refer to as `alternative clusters data set').

\begin{table}[h]
\small
\begin{center}
\begin{tabular}{cccc}
\hline
Type 	& $\alpha$	& $\beta$	&  Measurement\\
\hline
X-ray masses \citep{Vikhlinin09}	& 0.25	& 0.47	& $0.813\pm 0.013$\\
SZ \citep{Planck13_SZ}	& 0.27	& 0.301	& $0.782\pm 0.01$\\
\hline
X-ray luminosities \citep{Mantz10}	& 0.3	& 0.25	& $0.8\pm 0.04$\\
X-ray cross CMB \citep{Hajian13} & 0.3	& 0.26	& $0.8\pm 0.02$\\
SZ  \citep{Benson13}	& 0.25	& 0.298	& $0.785\pm 0.037$\\
X-ray counts \citep{Henry09} & 0.32	& 0.3	& $0.86\pm 0.04$ \\
Number counts \citep{Rozo10} & 0.25 & 0.41& $0.832\pm 0.033$\\
Number counts \citep{Tinker12} & 1.0 & 0.5 & $0.465\pm 0.03$ \\
\hline
\end{tabular}
\end{center}
\caption{\footnotesize
Cluster abundance measurements (in the form of $\Omega_{\rm M}^\beta\sigma_8$ combination)  included in our analysis, specifying the type of measurement and the corresponding values of the parameters $\alpha$ and $\beta$, according to eq.\eqref{Omega_sigma8}. The first set (above the horizontal line) shows the `clusters data set' that we use in our analysis; these measurements are in tension with the   $\La$CDM prediction and their error-bars are small, thus  in a joint analysis they would dominate. The second set (below the horizontal line) is the one we refer to as `alternative clusters data set', which involves measurements showing less tension with $\La$CDM.}
\label{tab:clusters}
\end{table}

\subsection{Redshift Space Distortions}
 RSD measurements are sensitive to the parameter combination $f\sigma_8$, where $f= d\ln \delta /d\ln a$ is the growth rate, and are, at present, the most powerful way to measure the evolution of the growth of structures. That is why we use these measurements to constrain growth. We use data from 6dF \citep{Beutler12}, BOSS-CMASS \citep{Reid14}, WiggleZ \citep{Blake11} and the VIMOS Public Extragalactic Redshift Survey (VIPERS) \citep{Torre13}. The measurements used in this analysis are shown in the right panel of Figure \ref{fig:Preds_data} along with the $\La$CDM prediction from Planck 2015 and they are presented in table \ref{tab:RSD}, specifying the survey of origin and the corresponding effective redshift $z_{\rm eff}$.

\begin{table}[h]
\small
\begin{center}
\begin{tabular}{ccc}
\hline
Survey	& $z_{\rm eff}$	&  $f\sigma_8$\\
\hline
6dF \citep{Beutler12} 	& 0.067	& $0.423\pm 0.055$\\
WiggleZ \citep{Blake11}	& 0.22	& $0.42\pm 0.07$\\
WiggleZ \citep{Blake11}	& 0.41	& $0.45\pm 0.04$\\
BOSS-CMASS \citep{Reid14}	& 0.57	& $0.450\pm 0.11$\\
WiggleZ \citep{Blake11}	& 0.60	& $0.43\pm 0.04$\\
WiggleZ \citep{Blake11}	& 0.78	& $0.38\pm 0.04$\\
VIPERS \citep{Torre13}	& 0.80	& $0.47\pm 0.08$\\
\hline
\end{tabular}
\end{center}
\caption{\footnotesize
RSD measurements included in our analysis, specifying the survey that obtained each measurement and the corresponding effective redshift, $z_{\rm eff}$.}
\label{tab:RSD}
\end{table}

%% file: sec3.tex
\section{Methodology}\label{sec:method}
\subsection{Background, conventions  and notation}
As noted before, in General Relativity  the recovered values for any two (split) parameters (in growth and geometry) must be consistent.
The corresponding  meta-parameters have the labels `geom' and `growth'. We let them vary independently and be constrained by observational data: `geom' effective parameters fit data from probes related to the expansion history and `growth' effective parameters are determined by measurements of growth of structures. 

We adopt the $w$CDM model\footnote{A flat model where the dark energy equation of state parameter, $w$, is constant.}  to account for variations in the behavior of dark energy. Thus we consider the following  possibilities for splitting dark energy parameters: only $w$, only $\Omega_{\rm DE}$, or both simultaneously. In the latter case, the increased dimensionality of the parameter space introduces parameter degeneracies which will be further discussed in Sec. \ref{sec:discussion}.

The quantities we use to constrain the geometry parameters are mostly distances inferred from  BAO and SNeIa surveys, and also the high multipoles of the angular power spectra of the CMB. All these observables depend on  the radial comoving distance $\chi$, which in turn depends on the expansion history through the Hubble parameter $H$:
\beq
\chi (z)=\int^z_0 \frac{cdz^\prime}{H(z^\prime)}\,.
\label{Comoving_dist}
\eeq
$H$ depends on the density parameters of the different species, whose evolution is determined by their corresponding equation of state. In the case of probes which constrain the expansion history parameters (and assuming a split in $w$ and $\Omega_{\rm DE}$) this is given by :
\beq
H^2(z)=H_0^2\left(\Omega_{\rm M}^{\rm geom}(1+z)^3+\Omega_{\rm r}(1+z)^4+\Omega_{\rm DE}^{\rm geom}(1+z)^{3(1+w_{\rm geom})}\right)\,,
\label{Hubble_param}
\eeq
where  the subscript $\rm M$ corresponds to non relativistic matter and $\rm r$ to radiation, i.e., photons and relativistic matter. We explicitly specify the splitting in $\Omega_{\rm M}$ to preserve geometry\footnote{We have chosen to work with a flat cosmology,  where $\Omega_{\rm M}+\Omega_{\rm DE}=1$. Therefore  in the case of split parameters we must have $\Omega_{\rm M}^{\rm geom}+\Omega_{\rm DE}^{\rm geom}=1= \Omega_{\rm M}^{\rm growth}+\Omega_{\rm DE}^{\rm growth}$.  Since $\Omega_{\rm M}=\Omega_{\rm b}+\Omega_{\rm CDM}$ and given that there are strong constraints in $\Omega_{\rm b}$ from several observables, the split in $\Omega_{\rm DE}$ translates into a split in $\Omega_{\rm CDM}$ and not in $\Omega_{\rm b}$.}.

Within the GR framework, growth of structures is described by a set of Boltzmann-Einstein equations. Although we use the exact system of Boltzmann-Einstein equations, for the sake of illustration, we remind the reader of the expression for the over density  $\delta=\delta \rho/\rho$  evolution in the case of a single fluid,
\beq
\frac{d^2\delta}{d\ln a^2}+\left[\frac{d\ln H}{d\ln a}+2\right]\frac{d\delta}{d\ln a}-\frac{3\Omega_{\rm M} H_0^2}{2a^3H^2}\delta =0\,,
\label{pert}
\eeq
where the Hubble parameter in this case would be calculated with growth parameters.
We note that different choices could be made when splitting the growth and geometry parameters. Any of these choices would serve the purpose of testing for a deviation from the null hypothesis.

\subsection{Fitting procedure}
We use the public Boltzmann code CLASS \citep{Blas11} and the Monte Carlo public code Monte Python \citep{Audren13} to fit our models to the data sets discussed in  Sec. \ref{sec:Data}. We  modify the codes to include the split parameters, add the RSD and clusters likelihoods and use growth or expansion parameters for  the different data sets as  indicated in table~\ref{tab:DATA}. We choose the  Metropolis Hastings algorithm as our sampling method and run sixteen Monte Carlo Markov Chains for each model until the fundamental parameters reach a convergence parameter $R-1 < 0.03$, according to the Gelman-Rubin criterion \citep{Gelman92}. In some cases (see sec.\ref{sec:NewPhys}),  we add or remove some data sets  using importance sampling to explore their effect on the posterior distribution. 

The fundamental cosmological parameters in our analysis are:
\beq
\lbrace \Omega_{\rm  b} h^2, \Omega_{\rm CDM}^{\rm geom}h^2,\Omega_{\rm CDM}^{\rm growth}h^2, H_0, A_S, n_s, \tau_{\rm reio}, w_{\rm geom}, w_{\rm growth}\rbrace\,,
\label{Param_fund}
\eeq
where when we split only in $w$ we impose $\Omega_{\rm CDM}^{\rm geom}=\Omega_{\rm CDM}^{\rm growth}$ and when we split only in $\Omega_{\rm CDM}$ we impose  $w_{\rm geom}= w_{\rm growth}$. Here $\Omega_{\rm b} h^2$ denotes the physical density of baryons, $h$, the reduced Hubble constant $H_0$/(100 km s$^{-1}$ Mpc$^{-1}$),  $\Omega_{\rm CDM}$, the cold dark matter density parameter, $A_S$, the primordial power spectrum scalar amplitude at $k=0.05$Mpc$^{-1}$, $n_s$, its spectral slope, and $\tau_{\rm reio}$, the integrated optical depth to the last scattering surface. Probes  sensitive to the  growth of structure (expansion history) will  constrain the growth (geometry) meta-parameters. All probes are also,  as usual, sensitive to the other, standard, cosmological parameters. For example, the low multipoles of the temperature and polarization CMB power spectra constrain growth meta-parameters but, as usual,  are key to  constrain $\tau_{\rm reio}$ as well.

We define the rest of our parameters as in the `base model' of Planck 2015 \citep{Planck15_Param}, including two massless neutrinos and a massive one ($m_\nu=0.06$ eV) and assuming an effective neutrino number $N_{\rm eff}=3.046$. In Section \ref{sec:nuphys} we also allow $N_{\rm eff}$ or $m_\nu$ to vary as free parameters to consider neutrino physics beyond the standard model. We assume flat priors on all the parameters of our model, setting a lower limit of 0 in $A_S$, $n_s$, $\tau_{\rm reio}$ and, when it corresponds, in $N_{\rm eff}$ and $\sum m_\nu$. 

As it is customary, we consider that
each data set is independent of the rest, so the total likelihood of the combined cosmological probes is the product of the individual likelihoods. 
In principle there could be a potential correlation between BAO and RSD as they are measured from the same survey. Moreover,
 BAO is sensitive to the Alcock-Paczynski effect, which may mimic the anisotropic clustering in redshift space due to bulk large-scale flows \citep{Kaiser87}. Nevertheless, direct estimations from measurements in redshift surveys indicate that this degeneracy is not so severe \citep{Samushia14}, so we can  safely consider both measurements  as if they were independent.
 
 The interpretation of the  constraints and the confidence levels is done as standard. Unless otherwise stated, the reported  error-bars on the parameters are the 68.3\% Bayesian central  confidence intervals. High-significance limits are  hard to calculate due to the scarcity of samples in these regions.  In some cases we have to estimate the significance of the null hypothesis (e.g., $w_{\rm growth}= w_{\rm geom}$) in this regime; this is  especially  evident when  samples  are found  only on one side of the  null hypothesis. In this case we estimate the significance in two ways.  We report a lower limit for the significance of the tension as probability=$1/N$, where $N$ is the total number of points in the chain after removing the burn-in phase, which we then convert to number of $\sigma$ under the Gaussian approximation (e.g, 68.3\% corresponds to 1$\sigma$, 95.4\% to 2$\sigma$  etc.). This is indicated by a  superscript $*$. We also report the significance in number, $n$, of 1$\sigma$ intervals (e.g., if  for the mean value $\langle w_{\rm growth}-w_{\rm geom}\rangle = \langle \Delta w\rangle \neq 0$ and $\sigma$ is the size of the 68.3\% Bayesian central  confidence interval of $\Delta w$, we report  $n$ such that $\langle \Delta w\rangle =n\sigma$). This is reported in brackets, following the  previous estimate.

%% file: sec4.tex
\section{Results with the full data set}\label{sec:Results}
In presenting our results we focus on the split dark energy parameters; the constraints on the cosmological parameters are summarized in table \ref{tab:results}. 
In the plots, red (blue) contours/shaded regions correspond to analysis done including the Planck 2015 (Planck 2013) data, illustrating the dependence of the joint analysis on the data release.
In the text we focus on the   results Planck 2015 , but we highlight the cases where significant differences appear. 
We start by presenting the analysis with all the data sets i.e. CMB, SNeIa, BAO, RSD and the clusters data set. Later, in Sec.~\ref{sec:NewPhys}, we consider the cases where tension between the split parameters is found and investigate if a particular data set is driving this result by considering different data combinations.

\subsection{Splitting parameters in the base model}
\label{sec:splitbasemodel}
\begin{figure}[h]
\minipage{0.5\textwidth}
\begin{center}
\includegraphics[width=0.9\textwidth]{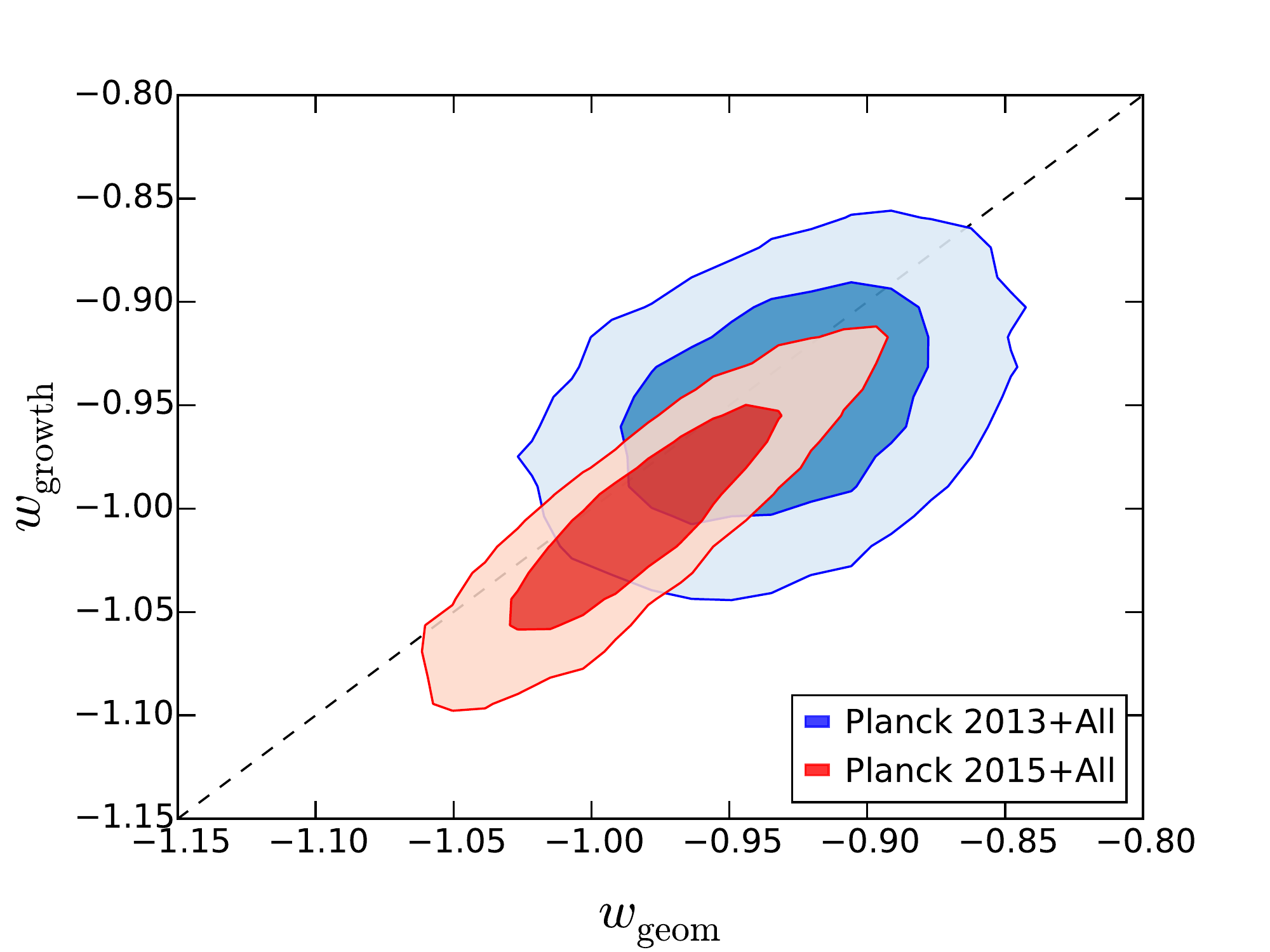}
\end{center}
\endminipage
\minipage{0.5\textwidth}
\begin{center}
\includegraphics[width=0.9\textwidth]{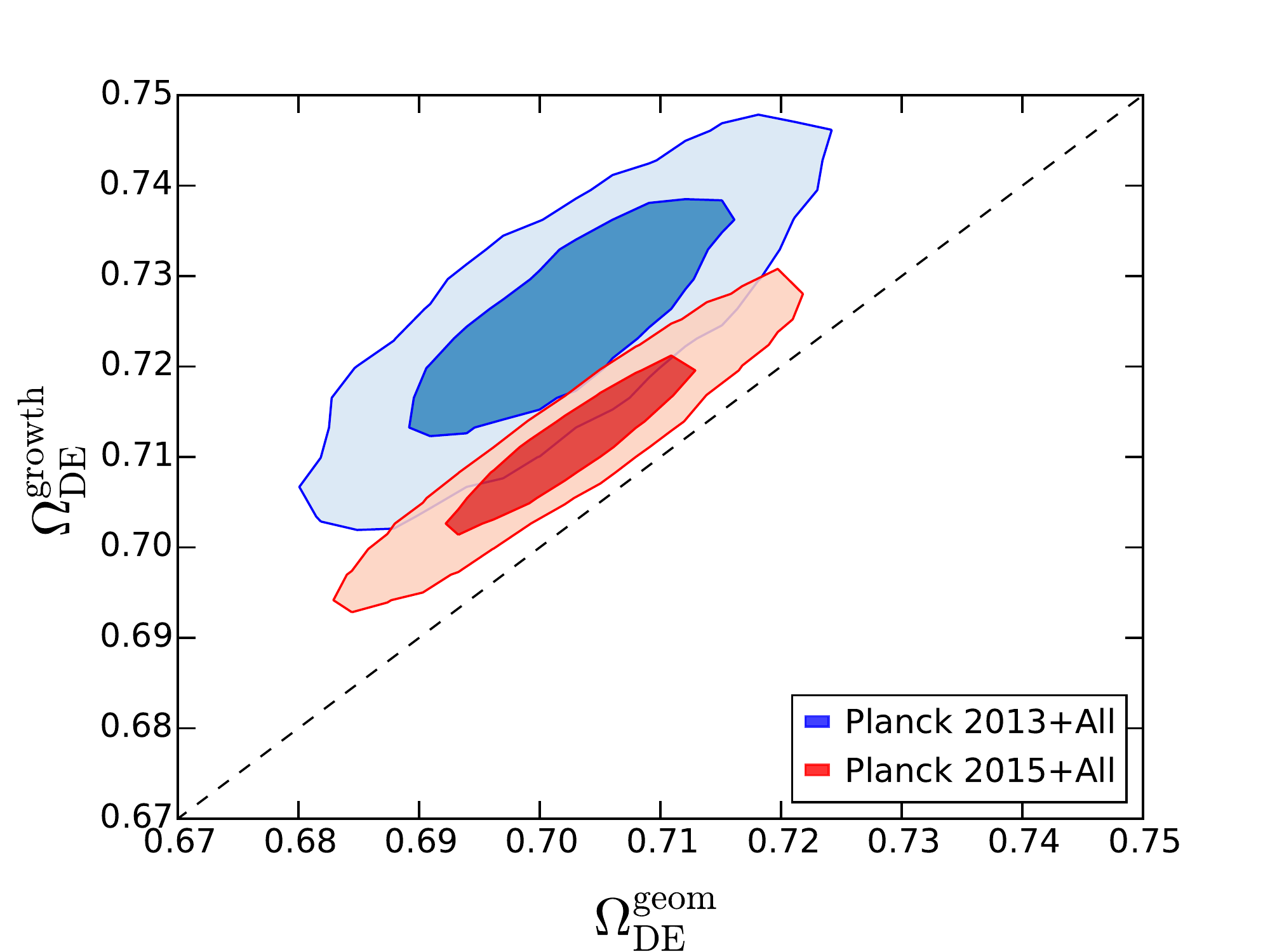}
\end{center}
\endminipage\hfill
\minipage{0.5\textwidth}
\begin{center}
\includegraphics[width=0.9\textwidth]{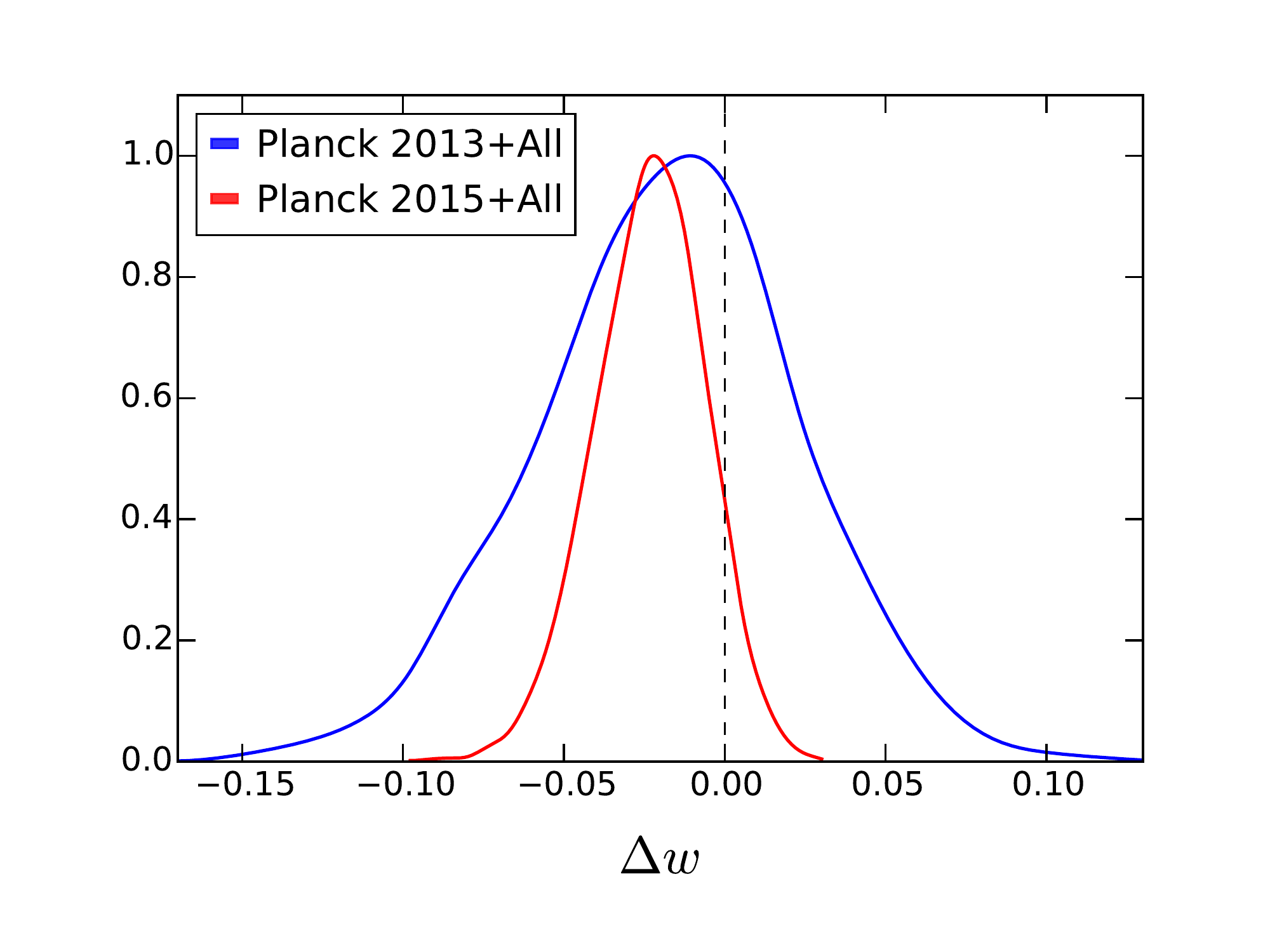}
\end{center}
\endminipage
\minipage{0.5\textwidth}
\begin{center}
\includegraphics[width=0.9\textwidth]{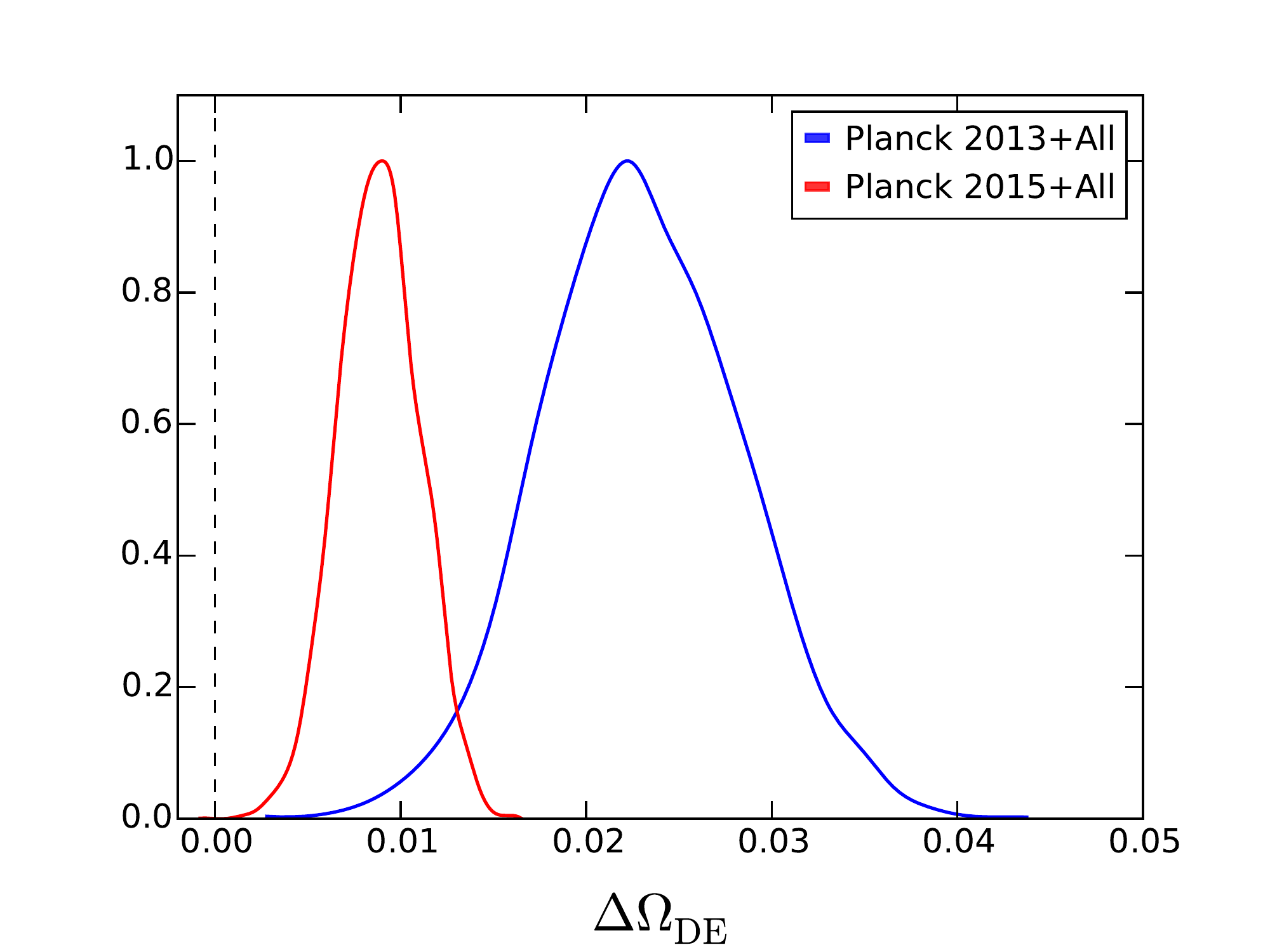}
\end{center}
\endminipage\hfill
\renewcommand{\baselinestretch}{1}
\caption{\footnotesize
\textit{Top}: Marginalized 68$\%$ and $95\%$ confidence level constraints in the $w_{\rm geom}$-$w_{\rm growth}$ plane in the case of only splitting the equation of state (\textit{left}) and in the $\Omega_{\rm DE}^{\rm geom}$-$\Omega_{\rm DE}^{\rm growth}$ plane in the case of only splitting the dark energy density parameter (\textit{right}). \textit{Bottom}: corresponding  distributions of $\Delta w$ and $\Delta \Omega_{\rm DE}$ in the case of only splitting the equation of state (\textit{left}) and in the case of only splitting the dark energy density parameter (\textit{right}). With ``All" we refer to the data sets from all the cosmological probes we use but CMB (i.e. RSD, BAO, SNeIa and cluster abundance).
The black dashed line indicates the points where the split parameters are equal. The results of the joint analysis using Planck 2013 data+All are shown in blue and the results using Planck 2015+All in red. 
}
\label{fig:w_split/Omega_split}
\end{figure}

We first split the dark energy parameters  starting from a base, flat wCDM model.
In the case where the equation of state of dark energy is split, marginalizing over all other parameters, we find $w_{\rm geom}=-0.98\pm 0.03$ and $w_{\rm growth}=-1.01\pm 0.04$ (both of them consistent with $w=-1$) and $\Omega_{\rm DE}=0.704\pm 0.008$. The constraints in the $w_{\rm geom}$-$w_{\rm growth}$ plane are shown in the top left panel of Figure \ref{fig:w_split/Omega_split} and the distribution of $\Delta w\equiv w_{\rm growth}-w_{\rm geom}$ is shown in the bottom left panel of Figure \ref{fig:w_split/Omega_split}
: the two parameters are nicely consistent. Note that using Planck 2013 data we find that  $w_{\rm geom}$ is still consistent with $w_{\rm growth}$ ($w_{\rm geom}=-0.93\pm 0.04$ and $w_{\rm growth}=-0.95\pm 0.04$), for a very similar dark energy density parameter: $\Omega_{\rm DE}=0.707\pm 0.008$. 
Thus splitting only $w$ shows no evidence of inconsistencies of General Relativity or unaccounted systematics in the data.

Splitting only in $\Omega_{\rm DE}$ we find that the equation of state is compatible with $\La$CDM ($w=-1.01\pm 0.03$) and that the split density parameters, marginalizing over all other parameters, are $\Omega_{\rm DE}^{\rm geom}=~0.703\pm 0.008$ and $\Omega_{\rm DE}^{\rm growth}=0.711\pm 0.008$.  However comparing the marginal distribution of each of the two parameters is not particularly informative, as it can be appreciated in the top right panel of Figure \ref{fig:w_split/Omega_split}. What is informative is the quantity $\Delta\Omega_{\rm DE}\equiv  \Omega_{\rm DE}^{\rm growth} - \Omega_{\rm DE}^{\rm geom}$, shown in the bottom right panel of Figure \ref{fig:w_split/Omega_split}. 
We find a tension of  $3.8\sigma$ between the two meta-parameters for $\Omega_{\rm DE}$ in the direction of  an excess of dark energy only felt by the growth of structures, which suppresses the clustering of large scale structures.

If we use the data of Planck 2013 instead of the data of Planck 2015, we find that the disagreement is larger, $\Omega_{\rm DE}^{\rm growth}$ increases to $0.725\pm 0.009$ and $\Omega_{\rm DE}^{\rm geom}$ does not change significantly, but the confidence region is wider (right panels of Figure \ref{fig:w_split/Omega_split}). In this case the tension is  $4.6\sigma$* (4.2$\sigma$).
 
\begin{figure}[h]
\minipage{0.5\textwidth}
\begin{center}
\includegraphics[width=0.9\textwidth]{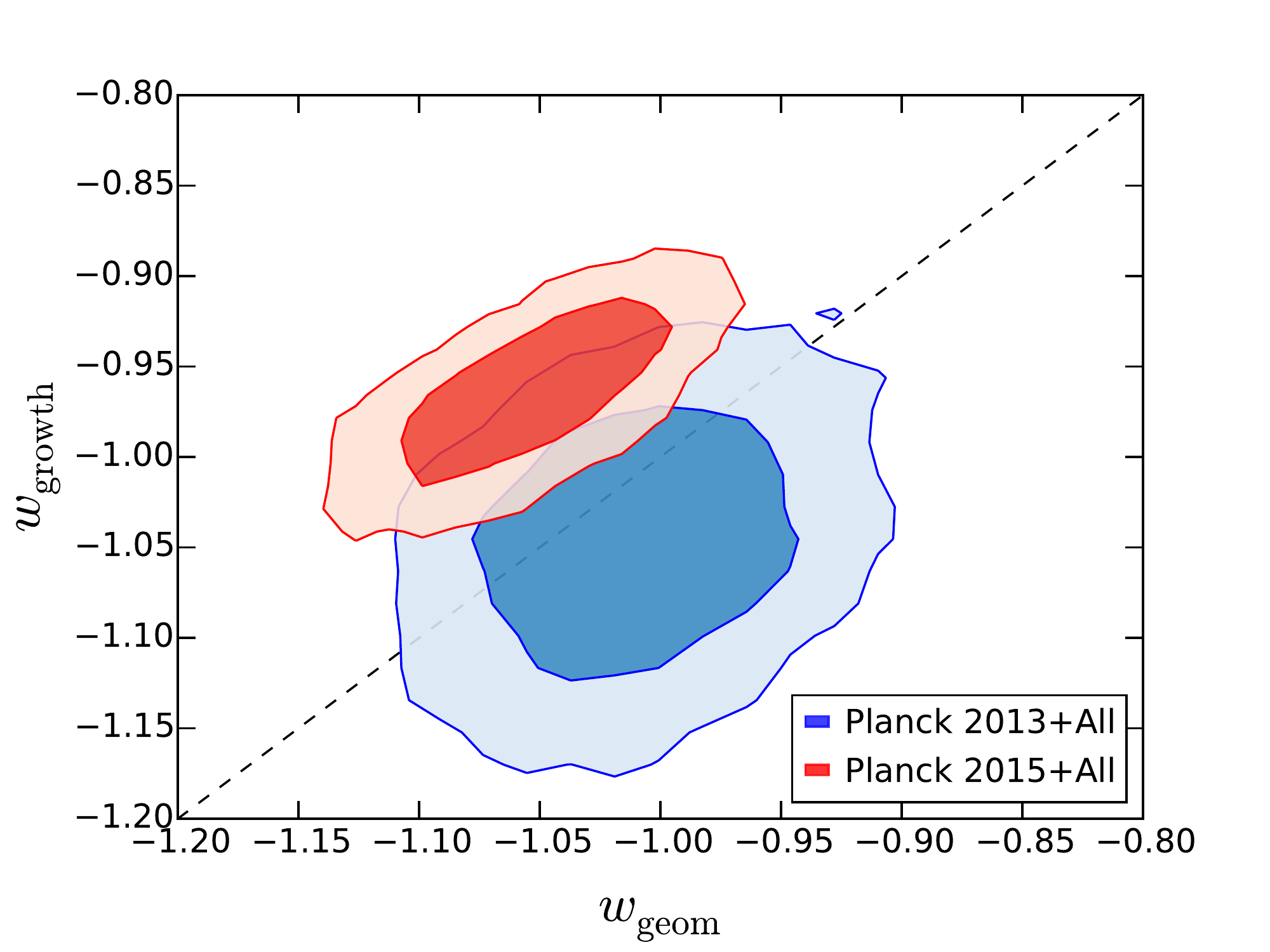}
\end{center}
\endminipage
\minipage{0.5\textwidth}
\begin{center}
\includegraphics[width=0.9\textwidth]{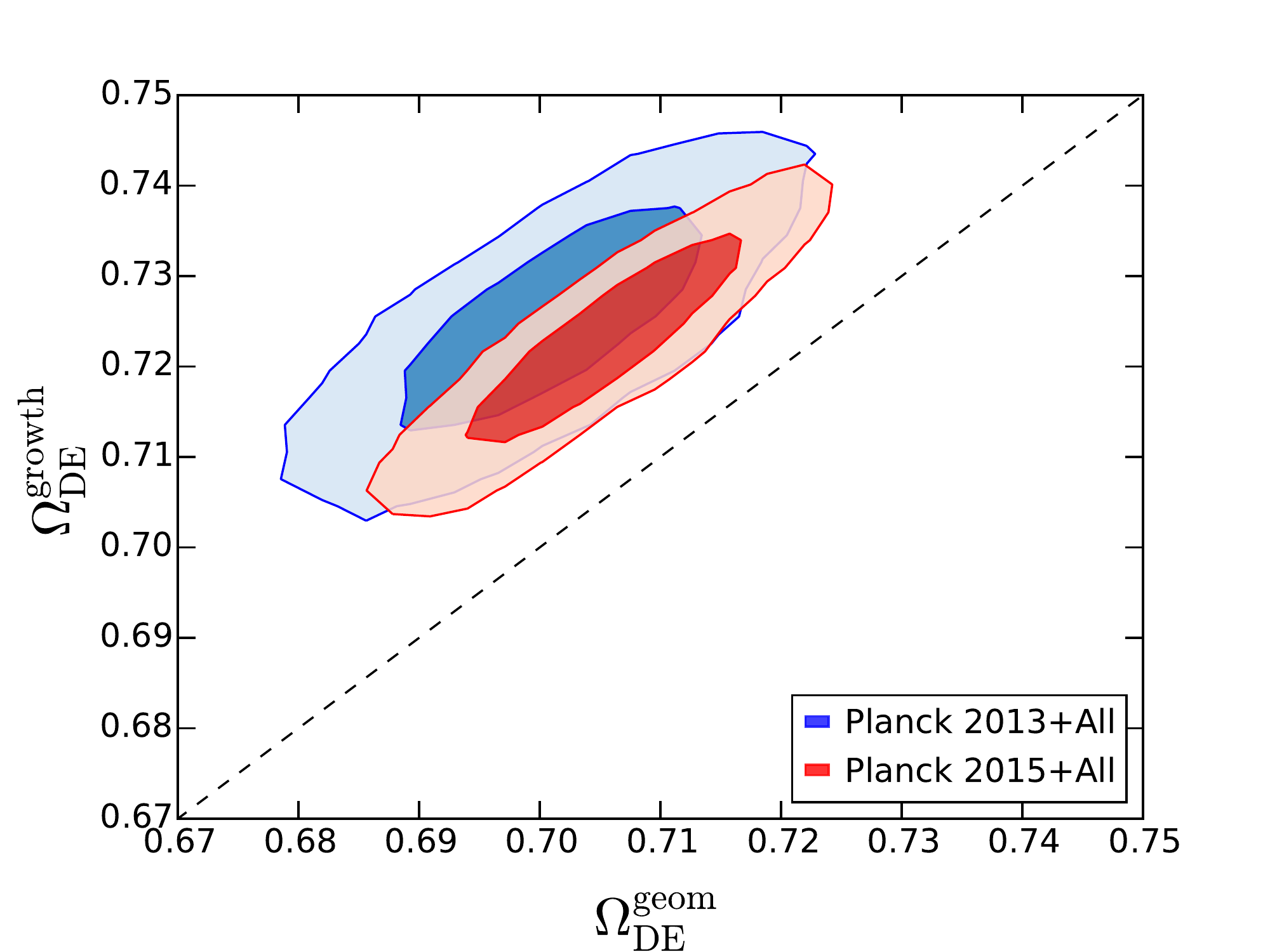}
\end{center}
\endminipage\hfill
\minipage{0.5\textwidth}
\begin{center}
\includegraphics[width=0.9\textwidth]{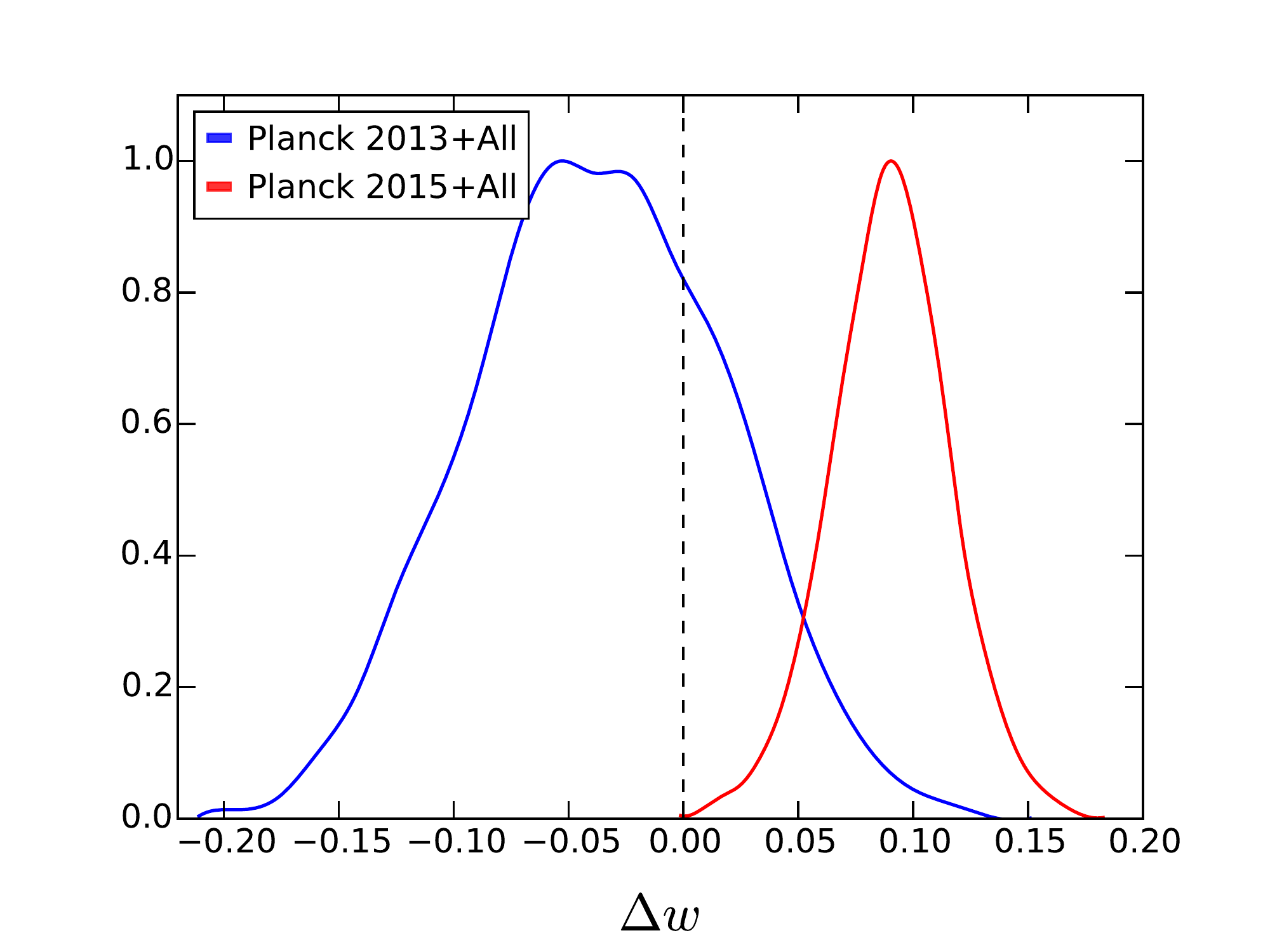}
\end{center}
\endminipage
\minipage{0.5\textwidth}
\begin{center}
\includegraphics[width=0.9\textwidth]{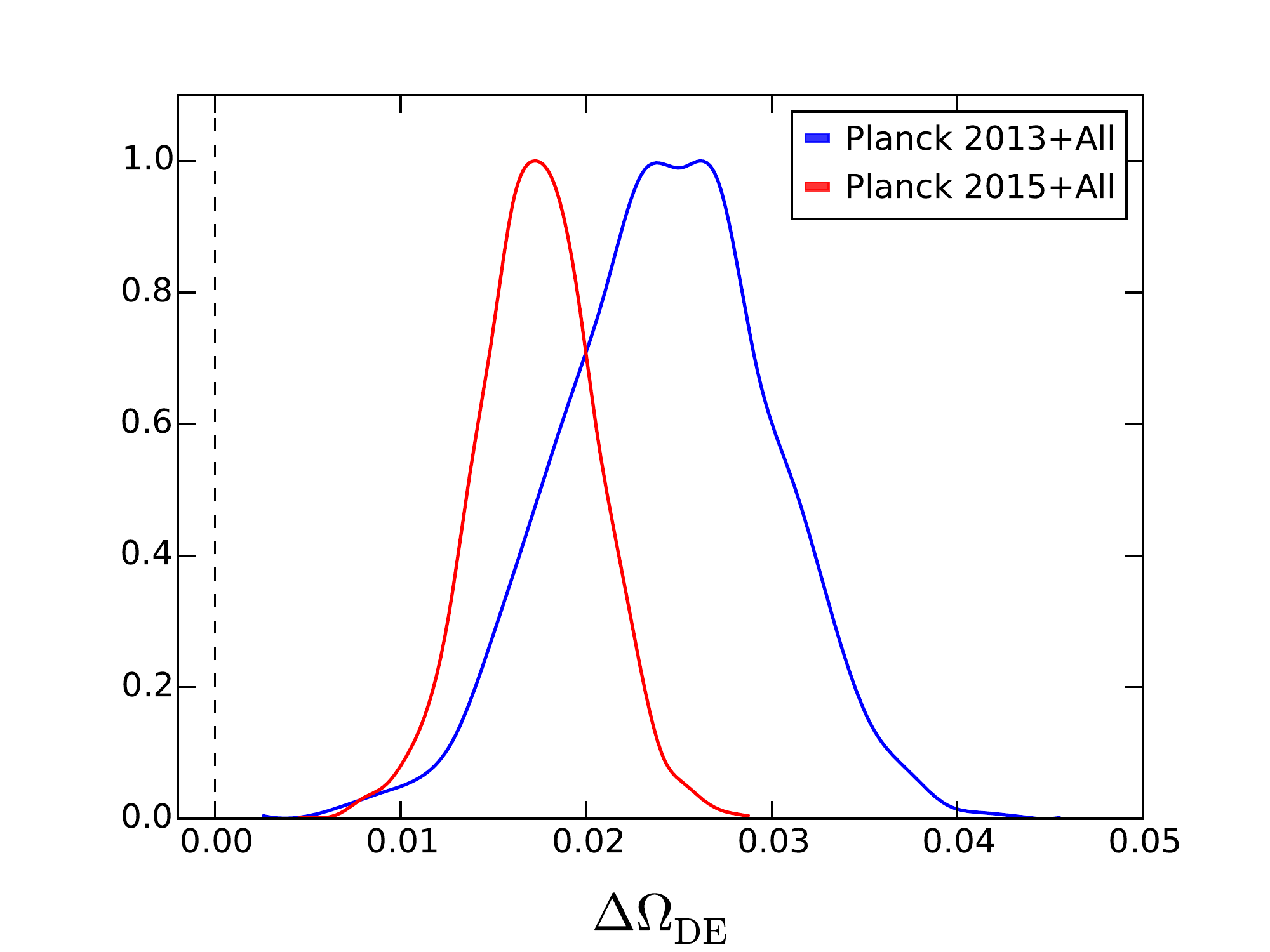}
\end{center}
\endminipage\hfill
\begin{center}
\minipage{0.5\textwidth}
\begin{center}
\includegraphics[width=0.9\textwidth]{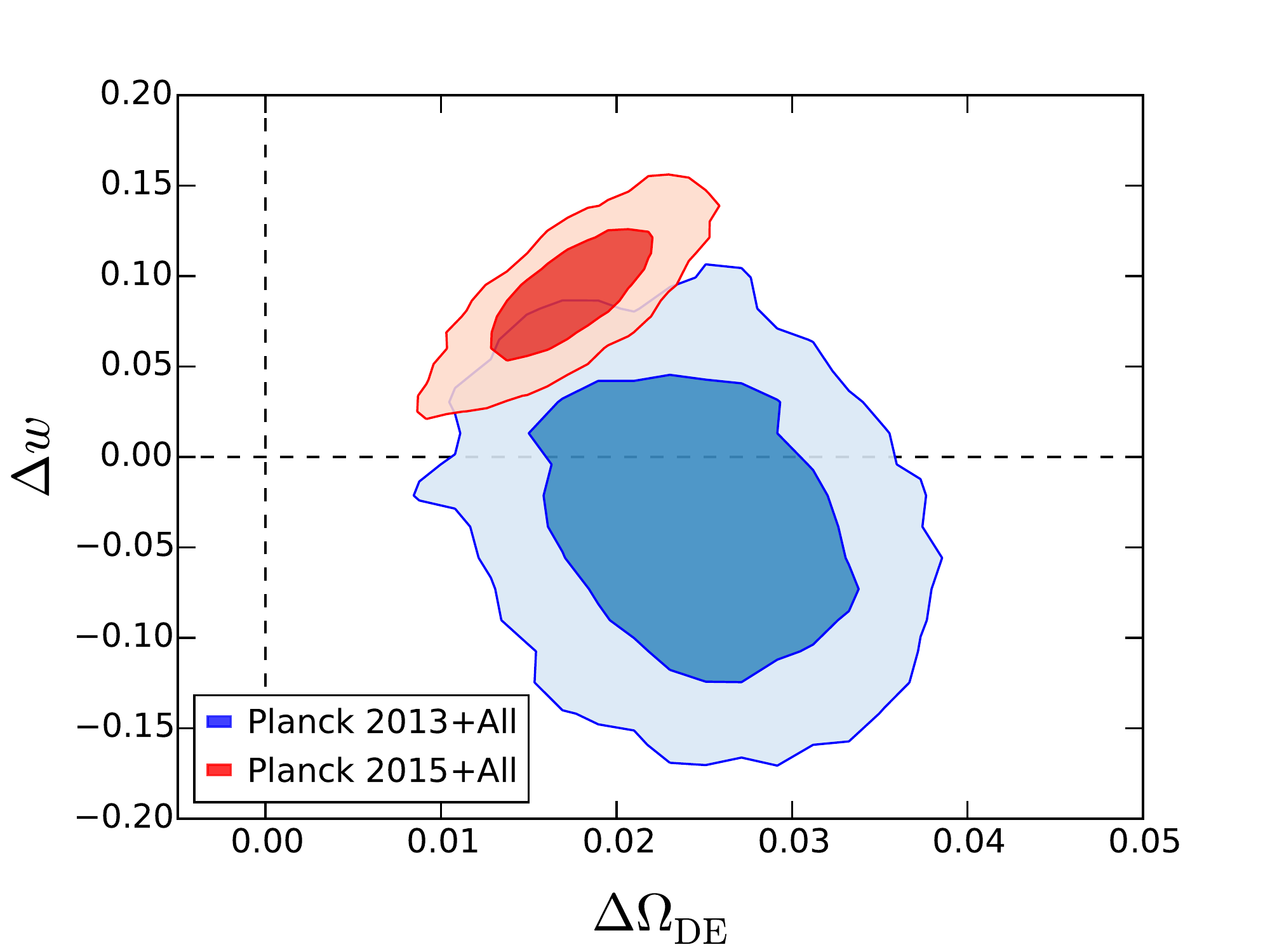}
\end{center}
\endminipage
\end{center}
\hfill
\renewcommand{\baselinestretch}{1}
\caption{\footnotesize
\textit{Top}: Marginalized 68$\%$ and $95\%$  confidence level regions in the $w_{\rm geom}$-$w_{\rm growth}$ plane (\textit{left}) and $\Omega_{\rm DE}^{\rm geom}$-$\Omega_{\rm DE}^{\rm growth}$ plane (\textit{right}) in the case of a simultaneous split in both dark energy parameters. \textit{Center}: corresponding distributions of $\Delta w$ (\textit{left}) and $\Delta \Omega_{\rm DE}$ (\textit{right}) in the case of splitting in $w$ and $\Omega_{\rm DE}$ simultaneously. \textit{Bottom}: Marginalized 68$\%$ and 95$\%$ confidence level regions in the $\Delta \Omega_{\rm DE}$-$\Delta w$ plane.
With ``All" we refer to the data sets from all the cosmological probes we use but CMB (i.e. RSD, BAO, SNeIa and cluster abundance).
The black dashed line indicates the points where the split parameters are equal. The results of the joint analysis using Planck 2013 data+All are shown in blue and the results using Planck 2015+All in red. 
}
\label{fig:double_split_constraints}
\end{figure}

When we split both $w$ and $\Omega_{\rm DE}$ at the same time we find  the posterior distributions (marginalized over all other parameters) reported in Figure \ref{fig:double_split_constraints}. The marginalized constraints on $w$ split parameters (top left panel of Figure \ref{fig:double_split_constraints}) are $w_{\rm geom}=-1.05\pm 0.04$ and $w_{\rm growth}=-0.96\pm 0.03$. 
This results are consistent with the fiducial value $w=-1$ but they present a tension of $3.5\sigma$, as it can be seen in the  left panels of the top and central sections of  Figure \ref{fig:double_split_constraints}. 

The  marginalized constraints on the dark energy density parameters (top right panel of Figure \ref{fig:double_split_constraints}) are $\Omega_{\rm DE}^{\rm geom}=0.705 \pm 0.008$ and $\Omega_{\rm DE}^{\rm growth}=0.722\pm 0.008$, but  (as the central right panel of Figure  \ref{fig:double_split_constraints} shows) the null hypothesis is excluded at   $4.4\sigma$* (5.3$\sigma$). The joint distribution of $\Delta w$ and $\Delta \Omega_{\rm DE}$ is shown in the bottom panel of Figure \ref{fig:double_split_constraints}.

Surprisingly (see Sec. \ref{sec:discussion}), using instead Planck 2013 data, the tension in $w$ disappears (the one-parameter marginalized constraints are $w_{\rm geom}=-1.01 \pm 0.04$ and $w_{\rm growth}=-1.05\pm 0.05$ and there is no evidence for deviations from $w_{\rm geom}=w_{\rm growth}$).  The tension in $\Omega_{\rm DE}$ is also lower in this case, but still very significant, 4.6$\sigma$* (4.2$\sigma$). As  before, the constraints are weaker using Planck 2013 data.

In the Appendix \ref{Appendix} we show the comparison of the data sets used in this paper with  predictions from some representative  best fit models. This is illustrated  in Figures \ref{fig:observables_geom} and \ref{fig:observables_growth}. Observables probing growth (expansion) are compared with best fit predictions for growth (geometry) parameters. The best fit for a $\La$CDM model  where  growth and expansion meta-parameters are forced to be equal and $w=-1$ is also shown for comparison. For details see the figure caption.

\subsection{Considering neutrino physics beyond \texorpdfstring{$\La$CDM}{LCDM}}
\label{sec:nuphys}
We have found indications that  $\Omega_{\rm DE}^{\rm growth}$ is inconsistent with  $\Omega_{\rm DE}^{\rm geom}$.
These results might be a signature of a failure of  GR  but of  systematics in the data as well. The tension, however, might also be provoked by  a failure of the base model, i.e., any physics beyond the standard cosmological model that affects the growth of structures and expansion history in a different way. Non standard neutrino properties have been advocated before to alleviate a related issue: that the CMB-inferred value of the parameter $\sigma_8$  is higher than the value measured by e.g., clusters of galaxies, weak gravitational lensing  or RSD e.g., \cite{Battye14,Dvorkin14,Beutler14,Battye15,MacCrann15}.
We therefore explore whether including extra degrees of freedom in the neutrino sector is favored by the data and thus alleviates the tensions found. 

Concretely, we  allow either the effective  number of neutrino  species $N_{\rm eff}$ or  the sum of neutrino masses $\sum m_\nu$, to  vary as free parameter  (with a flat prior and a lower limit of 0 in both cases). We perform this analysis in the cases where tension was found. 

When we  allow  $N_{\rm eff}$ to vary, we do not find any significant difference with respect to the previous findings: the tensions are not alleviated, and the mean value of $N_{\rm eff}$ is consistent with the fiducial value of  3.046. 

\begin{figure}[h]
\minipage{0.33\textwidth}
\begin{center}
\includegraphics[width=\textwidth]{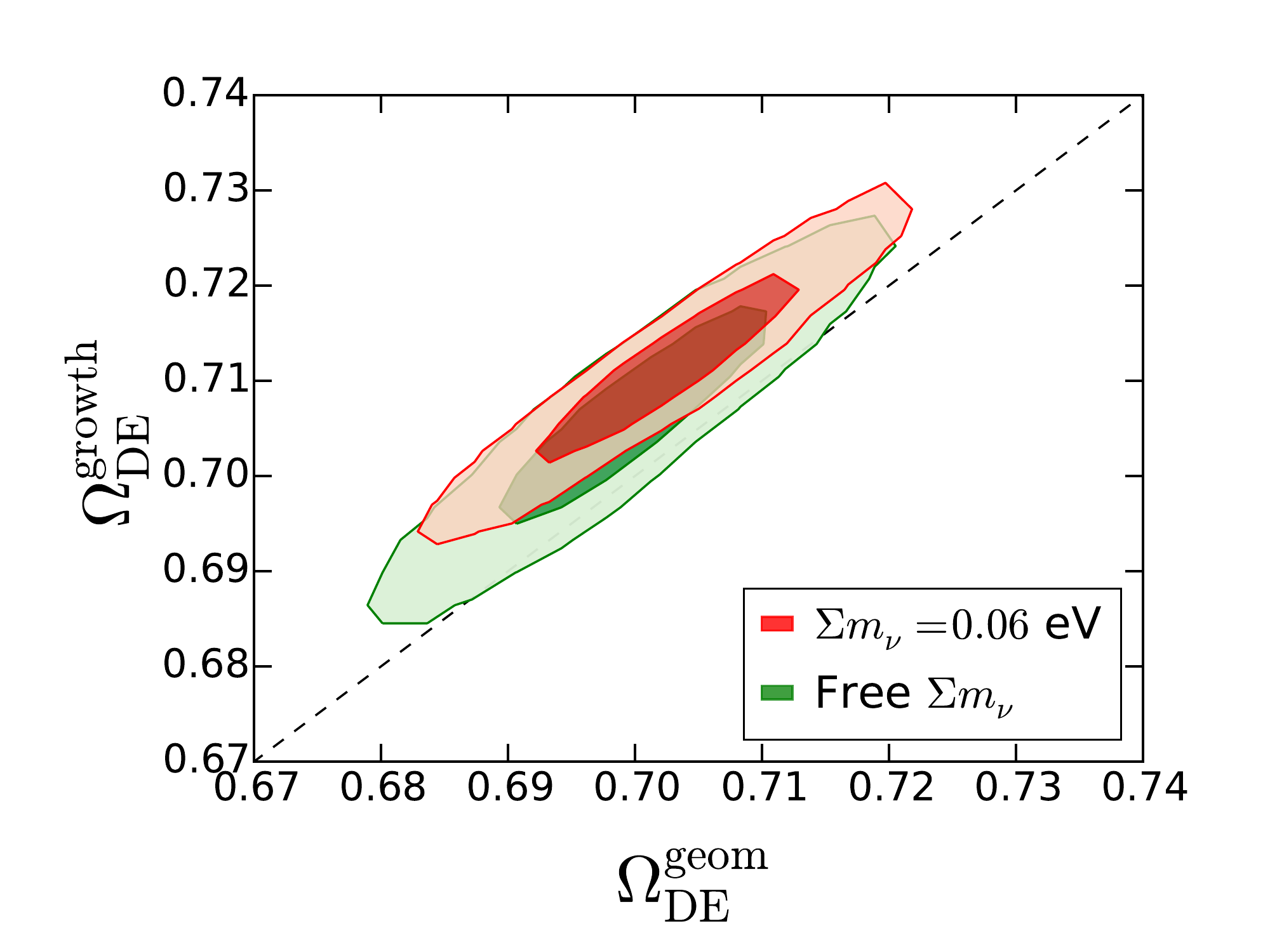}
\end{center}
\endminipage
\minipage{0.33\textwidth}
\begin{center}
\includegraphics[width=\textwidth]{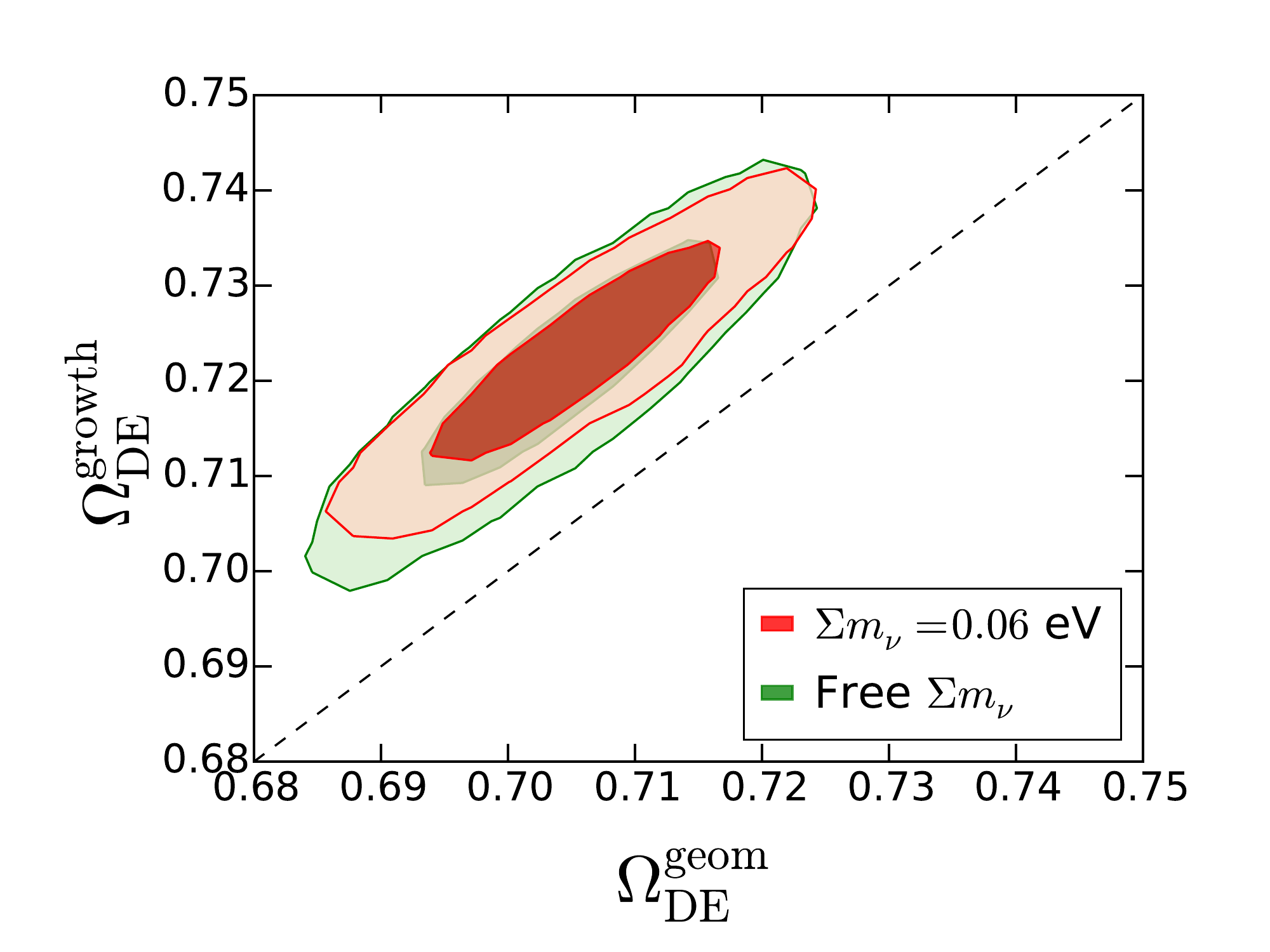}
\end{center}
\endminipage
\minipage{0.33\textwidth}
\begin{center}
\includegraphics[width=\textwidth]{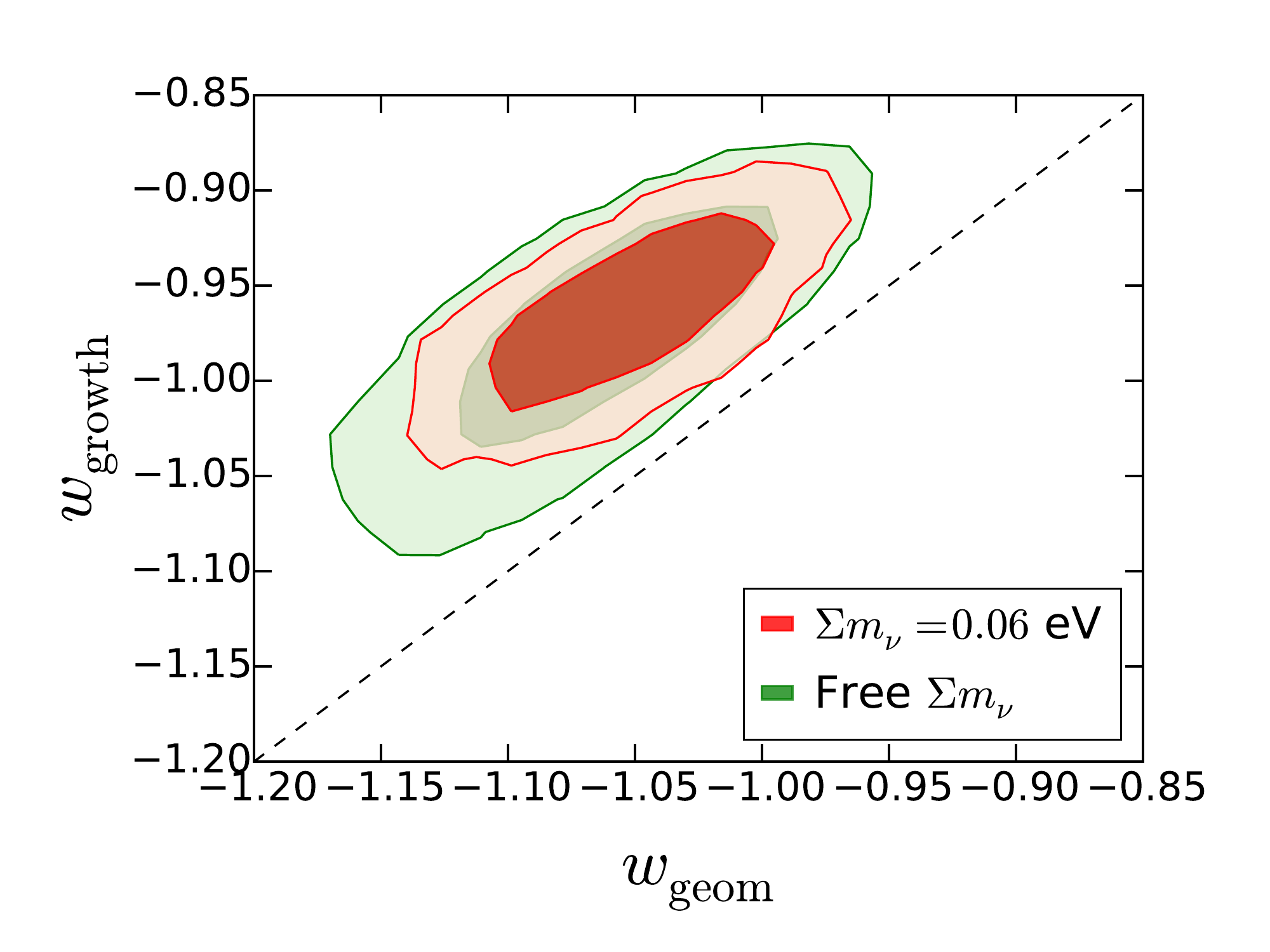}
\end{center}
\endminipage\hfill
\minipage{0.33\textwidth}
\begin{center}
\includegraphics[width=\textwidth]{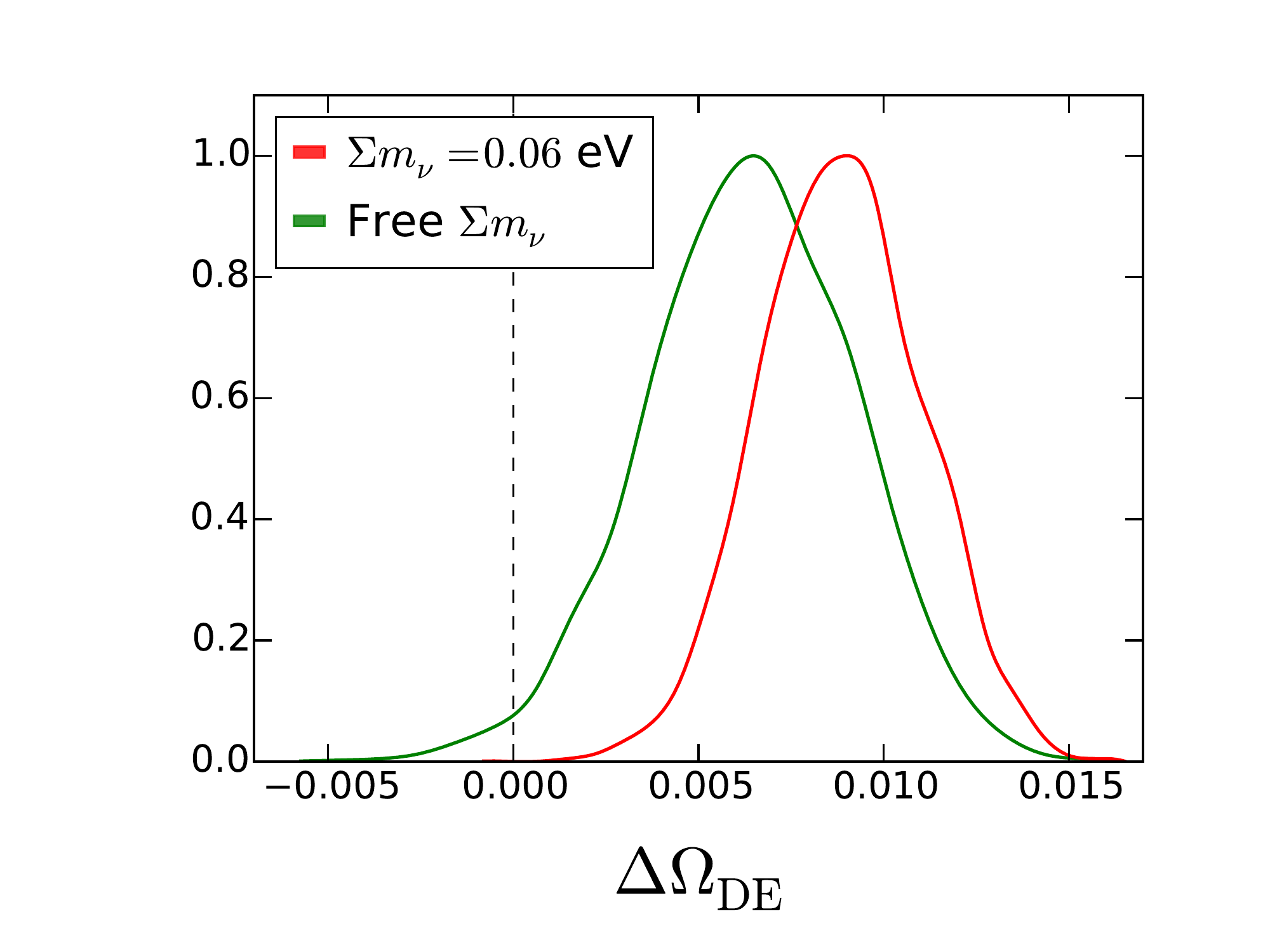}
\end{center}
\endminipage
\minipage{0.33\textwidth}
\begin{center}
\includegraphics[width=\textwidth]{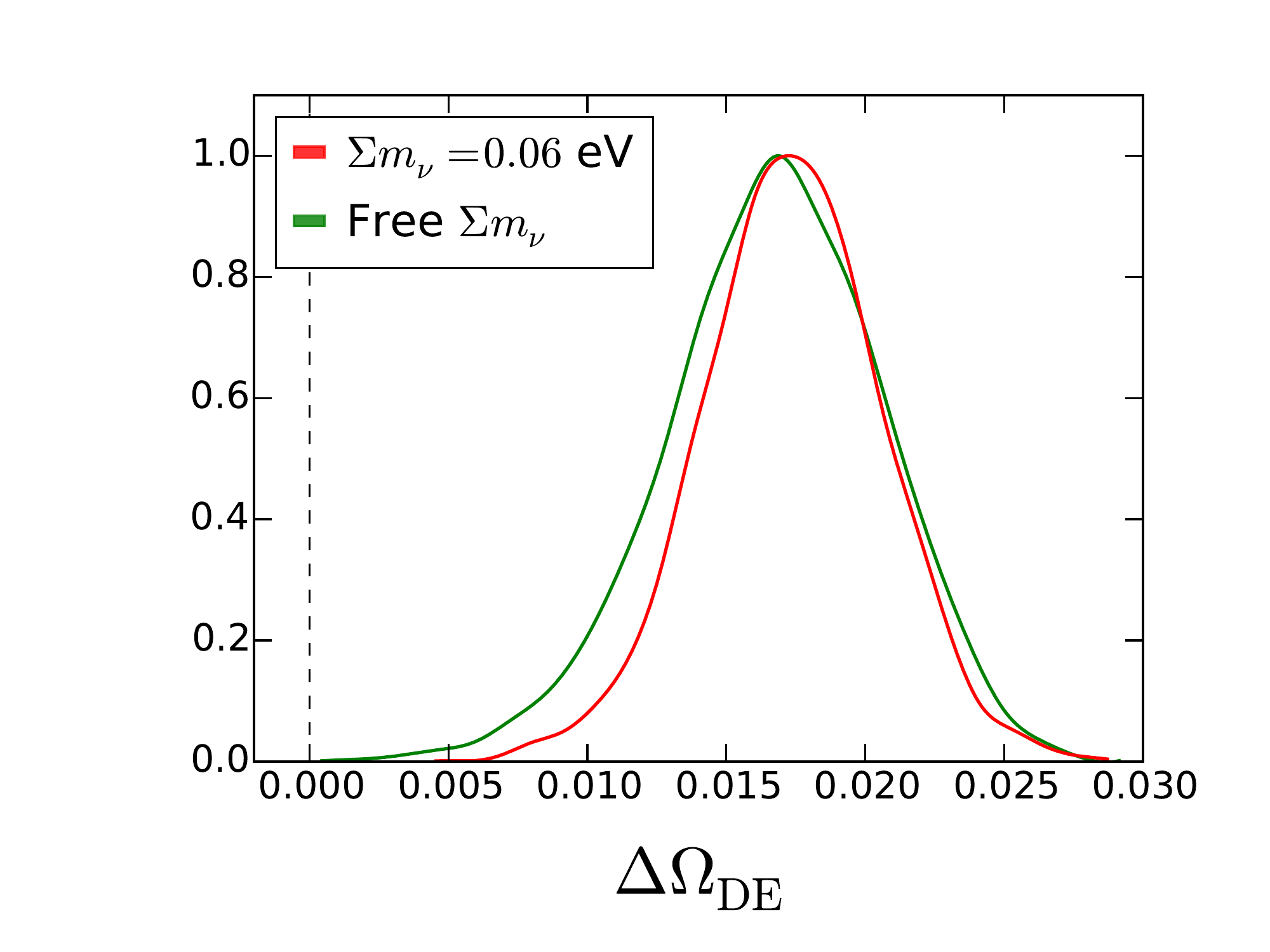}
\end{center}
\endminipage
\minipage{0.33\textwidth}
\begin{center}
\includegraphics[width=\textwidth]{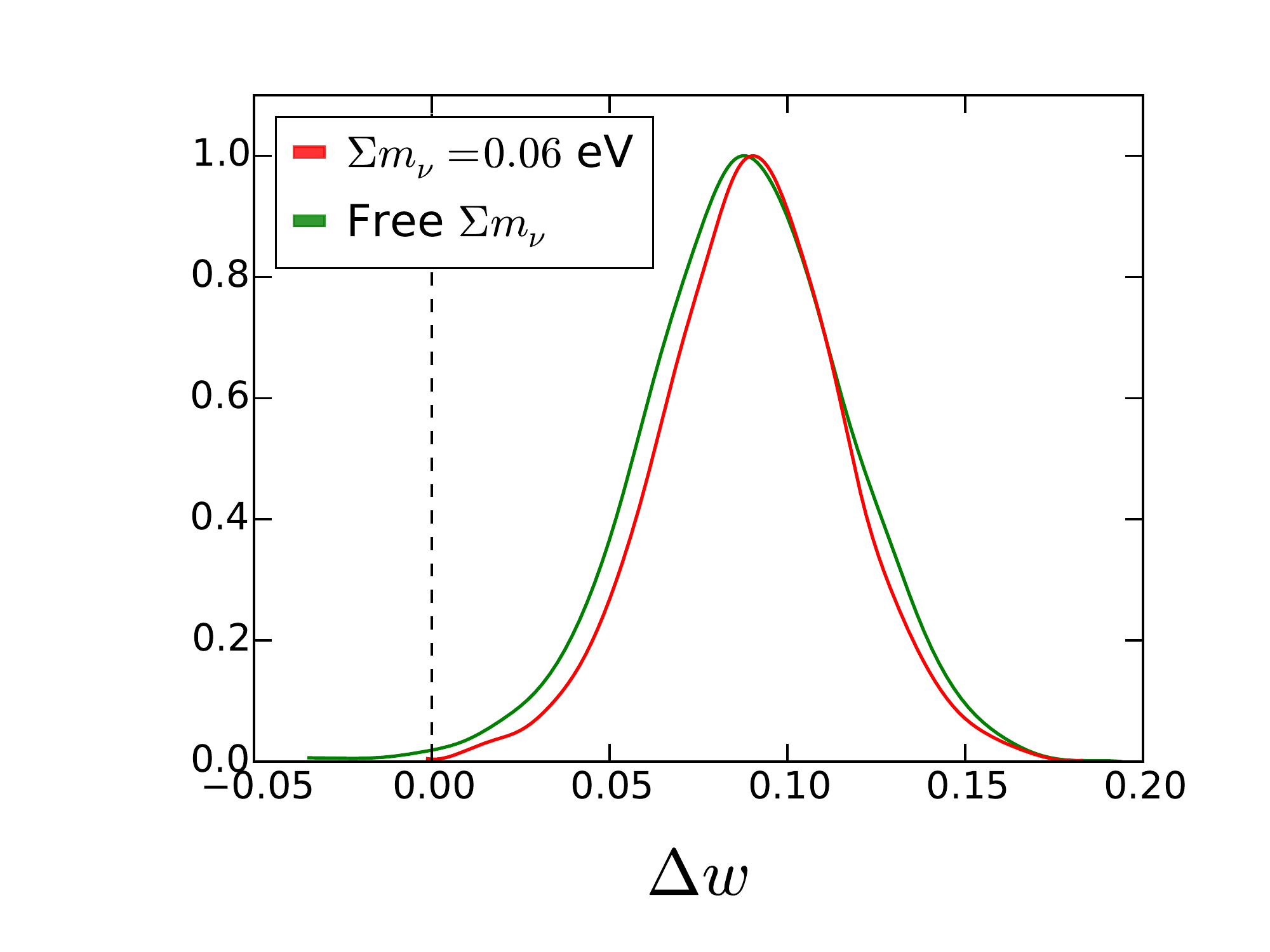}
\end{center}
\endminipage\hfill
\renewcommand{\baselinestretch}{1}
\caption{\footnotesize
\textit{Top}: 68$\%$ and $95\%$  confidence level  for marginalized constraints in the  $\Omega_{\rm DE}^{\rm geom}$-$\Omega_{\rm DE}^{\rm growth}$ plane and $w_{\rm geom}$~-~$w_{\rm growth}$ plane.  Left panel:  splitting only   $\Omega_{\rm DE}$. Middle and right panels:   splitting both $w$ and $\Omega_{\rm DE}$. \textit{Bottom}:
 corresponding distributions for $\Delta \Omega_{\rm DE}$ and $\Delta w$. The black dashed line indicates the points where the split parameters are equal. The red contours correspond to the results obtained with a fixed $\sum m_\nu=0.06$ eV (already shown in previous figures) and the green ones for $\sum m_\nu$ as a free parameter.}
\label{fig:neutrino}
\end{figure}

When $\sum m_\nu$ is left as a free parameter there are some quantitative, but not qualitative,  changes, which  are illustrated in Figure \ref{fig:neutrino}:  constraints  weaken slightly and tensions are reduced but  do not disappear. In particular, when  splitting only in $\Omega_{\rm DE}$, the disagreement is reduced from 3.8 to $2.4\sigma$ and we find the following constraints: $w=-1.06\pm 0.05$, $\Omega_{\rm DE}^{\rm geom}=0.700\pm 0.008$, $\Omega_{\rm DE}^{\rm growth}=0.706\pm 0.009$ and $\sum m_\nu=0.21\pm 0.10$ eV. When we split both $w$ and $\Omega_{\rm DE}$ the tension is reduced  from $3.5 \sigma$ and 4.4$\sigma$* ($5.3\sigma$) to  $2.8\sigma$  and  $4.1\sigma$ respectively, with $\sum m_\nu=0.087\pm 0.053$ eV. Using Planck 2013 data, the tensions are reduced in $\sim 1\sigma$ since the constraints are weaker.

\begin{table}
\scriptsize
\begin{center}
\makebox[\textwidth]{
\begin{tabular}{|c|c|c|c|c|c|c|}
 \hline
Parameter & $\La$CDM &$w$ split & $\Omega_{\rm DE}$ split & $w$ and $\Omega_{\rm DE}$ split 	& $\Omega_{\rm DE}$ split (free $\Sigma m_\nu$) & $w$ and $\Omega_{\rm DE}$ split (free $\Sigma m_\nu$)\\ \hline
$100\Omega_{\rm b} h^2$	& $2.253\pm 0.014$	& $2.252\pm 0.015$	& $2.238\pm 0.015$	& $2.233\pm 0.015$	& $2.238\pm 0.016$	& $2.231\pm 0.015$\\
$H_0$	&$69.0\pm 0.5$	 & $68.5\pm 1.0$	& $68.7 \pm 0.9$	& $69.3\pm 0.9$		& $68.8 \pm 0.9$	& $69.4\pm 1.0$\\
$10^9A_S$	& $2.05\pm 0.04$	&$2.02\pm 0.06$	& $2.05\pm 0.04$	& $2.19\pm 0.07$	& $2.12\pm 0.07$	& $2.20\pm 0.07$\\
$n_s$	& $0.971\pm 0.004$	&$0.971\pm 0.004$	& $0.969\pm 0.004$	& $0.967\pm 0.005$	& $0.969\pm 0.004$	& $0.967\pm 0.005$\\
$\tau_{\rm reio}$	& $0.047\pm 0.012$	&$0.039\pm 0.017$	& $ 0.046\pm 0.014$	& $0.077\pm 0.017$	& $ 0.063\pm 0.018$	& $0.079\pm 0.017$\\
$\Omega_{\rm DE}^{\rm geom}$	& \multirow{2}{*}{$0.708\pm 0.006$}		& \multirow{2}{*}{$0.704\pm 0.008$}	& $0.703\pm 0.008$	& $0.705\pm 0.008$	& $0.700\pm 0.008$	& $0.704\pm 0.008$\\
$\Omega_{\rm DE}^{\rm growth}$ &	&	& $0.711\pm 0.008$	& $0.722\pm 0.008$	& $0.706\pm 0.009$	& $0.721\pm 0.009$\\
$w_{\rm geom}$	&  \multirow{2}{*}{--}	&$-0.98\pm 0.03$	& \multirow{2}{*}{$-1.01\pm 0.03$}	& $-1.05\pm 0.04$		& \multirow{2}{*}{$-1.06\pm 0.05$}	& $-1.06\pm 0.04$\\
$w_{\rm growth}$ &	& $-1.01\pm 0.04$	&	& $-0.96\pm 0.03$		&	& $-0.97\pm 0.04$	\\
$\Sigma m_\nu$	(eV)&-- & -- & -- & -- & $0.21 \pm 0.10$ & $0.087 \pm 0.053$  \\
\hline
\end{tabular}}
\end{center}
\caption{\footnotesize
Constraints on the cosmological parameters from the joint analysis of all observational data sets included in our work. Each column corresponds to one model: $\La$CDM, $w$ split, $\Omega_{\rm DE}$ split and both of them split. The last two columns show the results for the last two models with the sum of neutrino masses is a free parameter.
In the cases when $w$ or $\Omega_{\rm DE}$ are not split, the constraints which are shown at a mid level correspond to the unsplit parameter.}
\label{tab:results}
\end{table}

%% file: sec5.tex
\section{New physics, inadequacy of the modeling or systematic errors?}\label{sec:NewPhys}
Before claiming  evidence of deviations from GR within a minimally coupled dark energy, it is necessary to rule out the rest of possible explanations for the tensions found. 
Since tensions arise mostly in the growth meta-parameters, any change in the amplitude of clustering would cause not just a change in the growth rate but also a change in the shape of the matter power spectrum. Current data is not constraining enough to allow significant freedom in the matter power spectrum shape,  use meta-parameters for dark energy, and still obtain useful constraints.
We have to limit investigation to specific models. In particular, since non-zero neutrino masses have been invoked in the literature before, this is the model we explore.

As seen in Sec. \ref{sec:nuphys}, considering non-standard neutrino properties reduces but does not eliminate the tensions. In the best case, when we split only $\Omega_{\rm DE}$ and allow $\sum m_{\nu}$ to vary, the tension reduces (just) to 2.4$\sigma$, but our value of $\sum m_\nu$ is inconsistent with other constraints in the literature which push the upper 95\% limit to $\sum m_\nu<0.12$ eV (e.g.,  \citep{WiggleZ, Palanque14, Palanque15, Viviinprep}).  This indicates that  massive neutrinos can not be the (full) solution to this problem, as already discussed e.g., in \cite{Leisted14}.

Another possibility is that 
dark energy requires a modeling more complex than a constant equation of state. In that case our $w$ values would be ``effective"   but, as long as the expansion history of the Universe determines uniquely the growth history, which is a feature of GR, the split parameters should still coincide if the  data used to constrain growth and geometry probe the same redshift range, as it is, at least approximately, in our case. 

Of course, tensions may arise because of 
unaccounted systematics or wrong modeling of the data. In principle, parameter splitting offers a powerful tool to test for it. With redundancy, when different data available probing the same quantity (e.g., growth) in the same redshift range and over the same scales, results from different data sets can be compared. This can be used to uncover ``the odd one out", likely affected by systematics. Conversely, if all data sets show tension on the same meta-parameter, that would be a hint of new physics which is not accounted for by the model.

With currently available data sets redundancy is limited. Yet, here we attempt to test for systematics in this way, even if only as a proof of principle. To explore this possibility we have performed the analysis removing and adding one by one growth data sets, using importance sampling whenever possible\footnote{We run chains without clusters and RSD data and then add them via importance sampling.}, in the cases where we found tensions. Both RSD and clusters data probe the growth of structures (the observable showing tension) roughly in the same redshift range.

\begin{figure}[h]
\minipage{0.33\textwidth}
\begin{center}
\includegraphics[width=\textwidth]{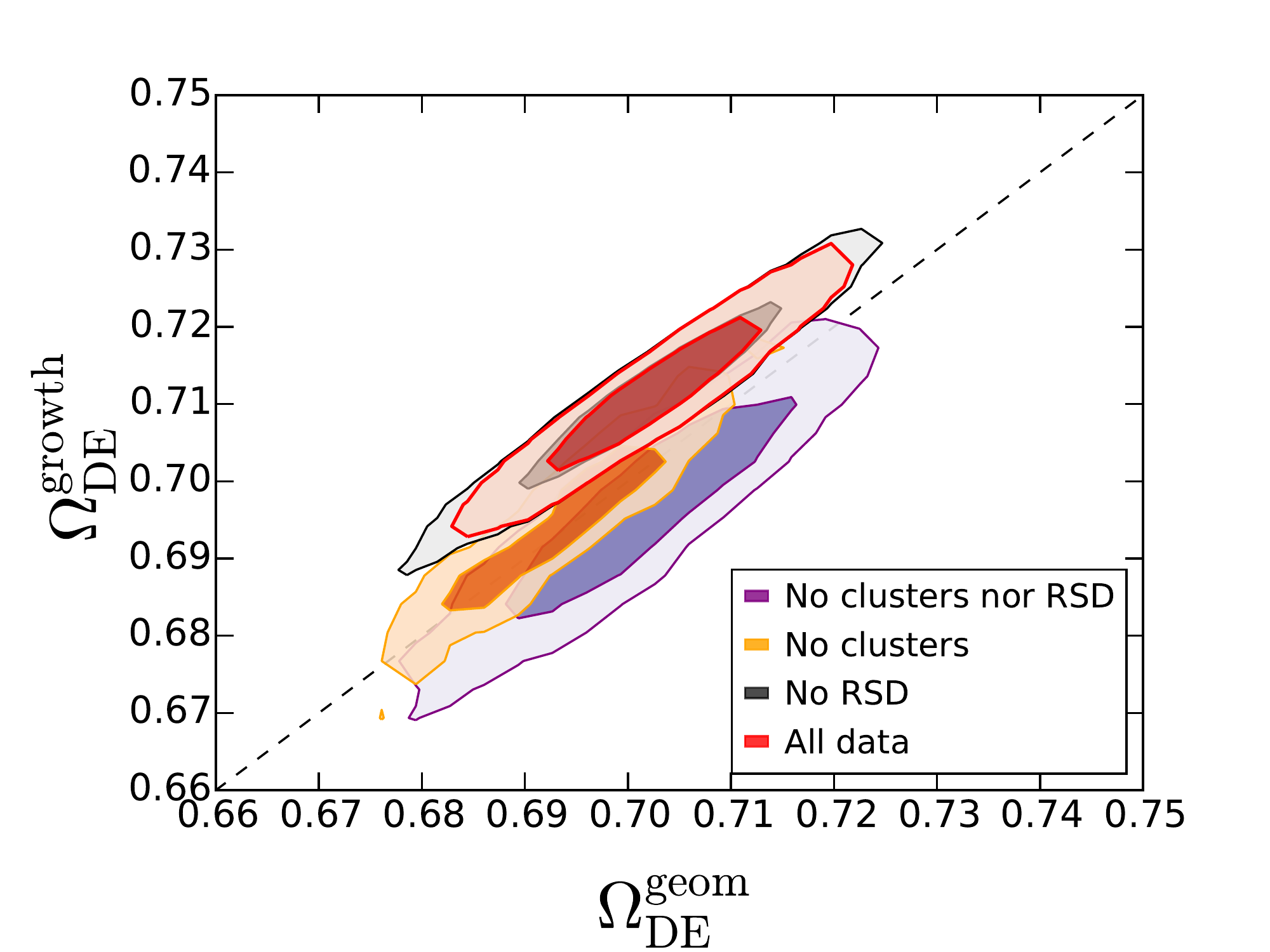}
\end{center}
\endminipage
\minipage{0.33\textwidth}
\begin{center}
\includegraphics[width=\textwidth]{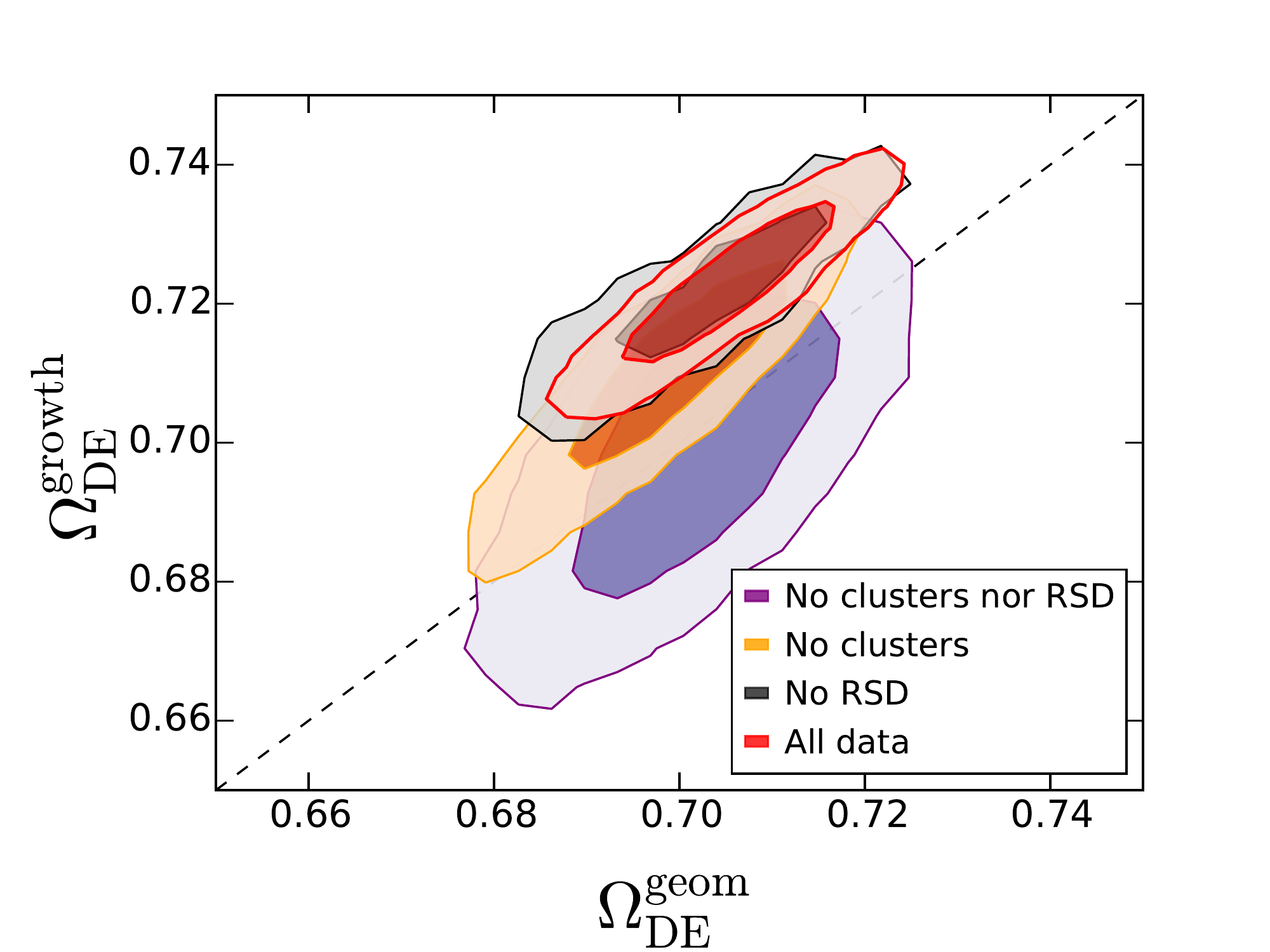}
\end{center}
\endminipage
\minipage{0.33\textwidth}
\begin{center}
\includegraphics[width=\textwidth]{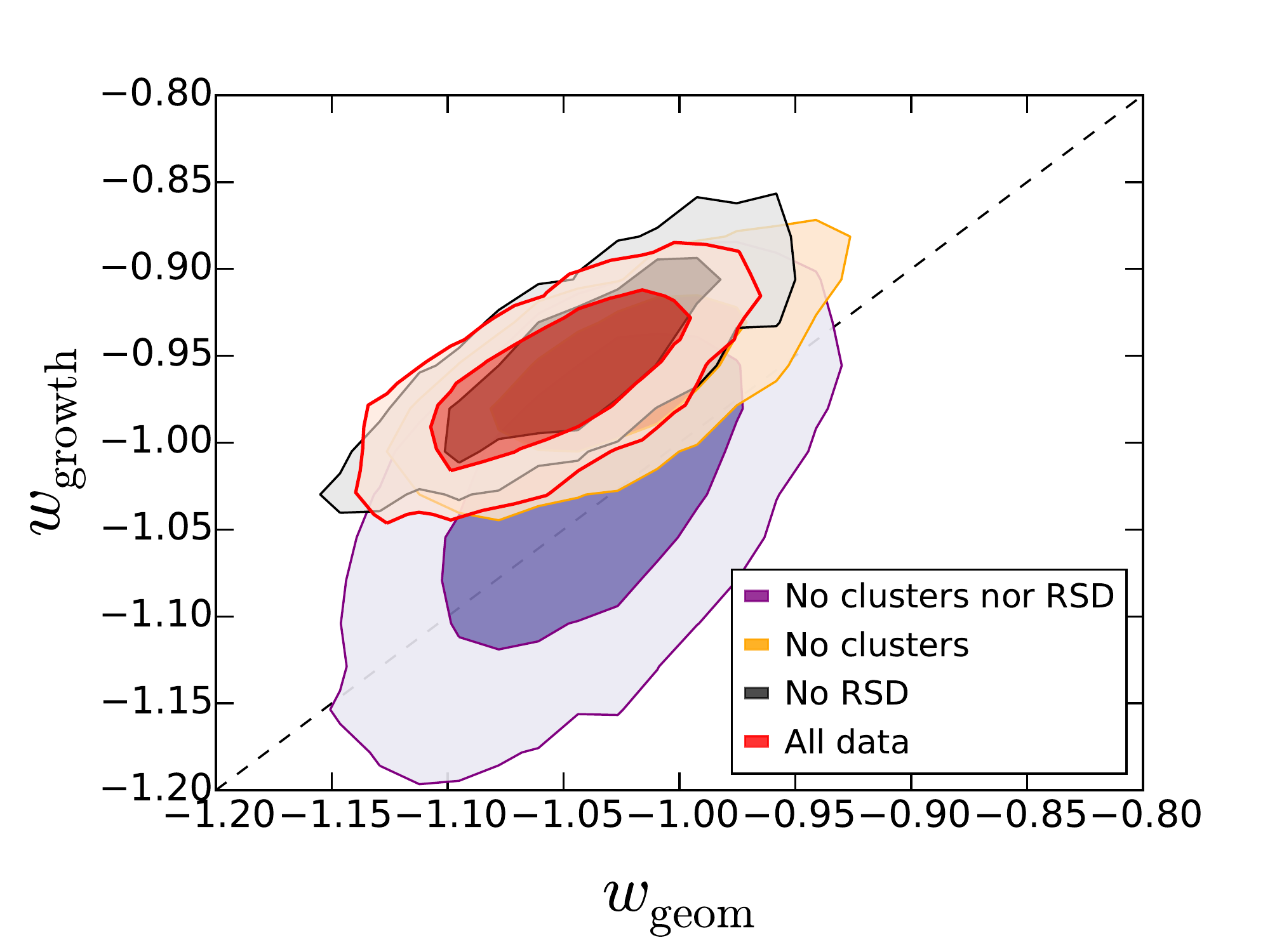}
\end{center}
\endminipage\hfill
\minipage{0.33\textwidth}
\begin{center}
\includegraphics[width=\textwidth]{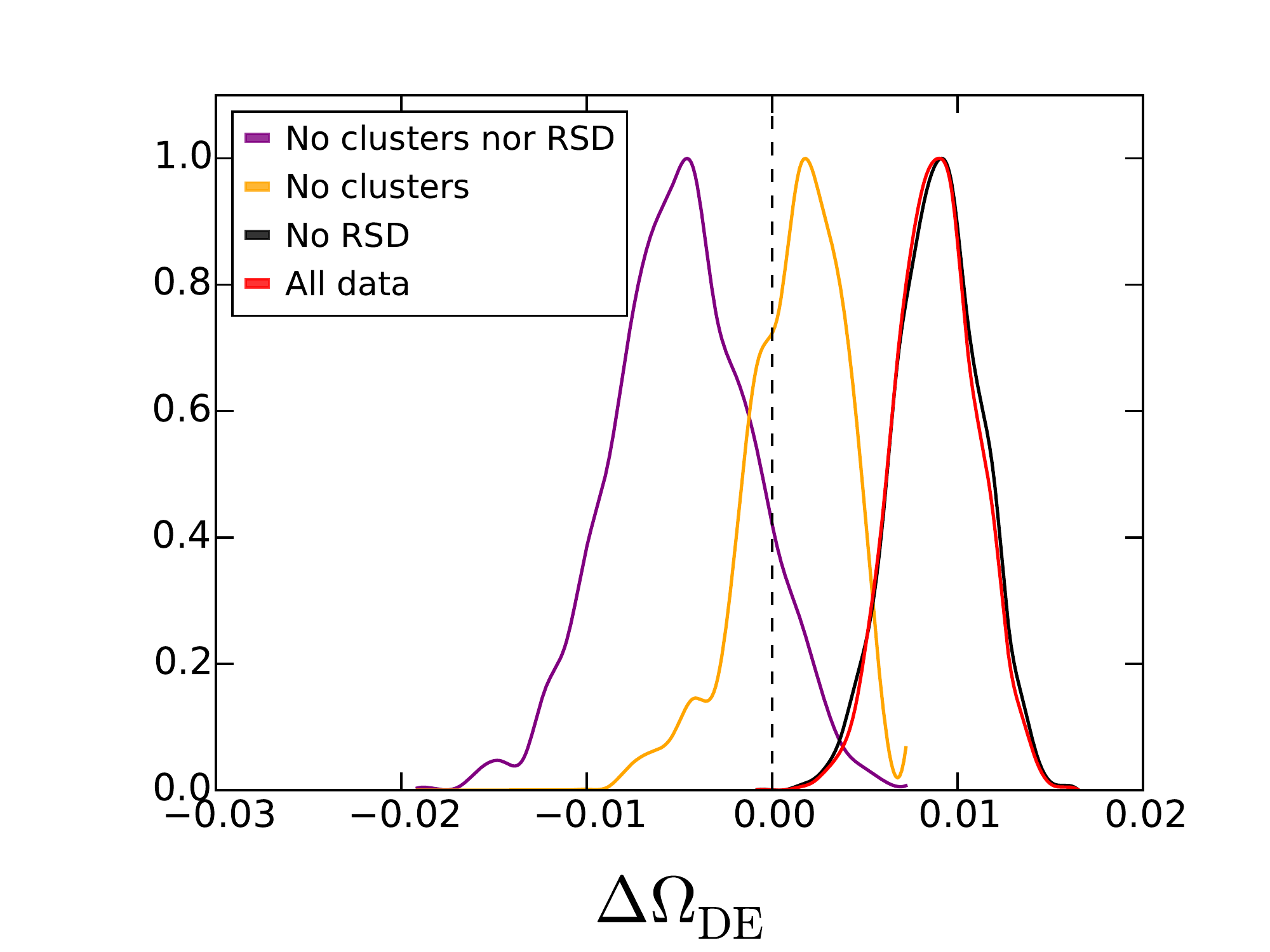}
\end{center}
\endminipage
\minipage{0.33\textwidth}
\begin{center}
\includegraphics[width=\textwidth]{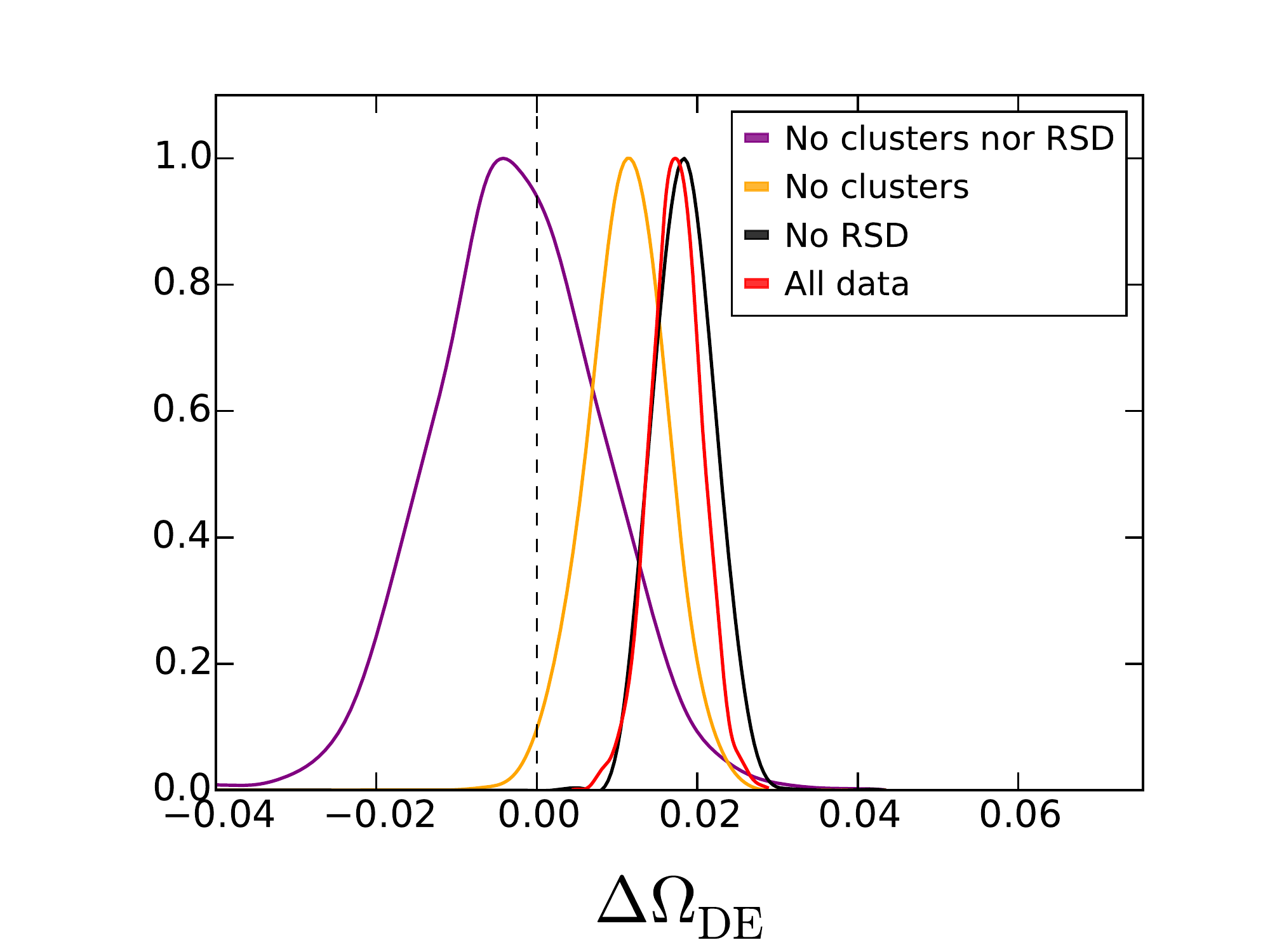}
\end{center}
\endminipage
\minipage{0.33\textwidth}
\begin{center}
\includegraphics[width=\textwidth]{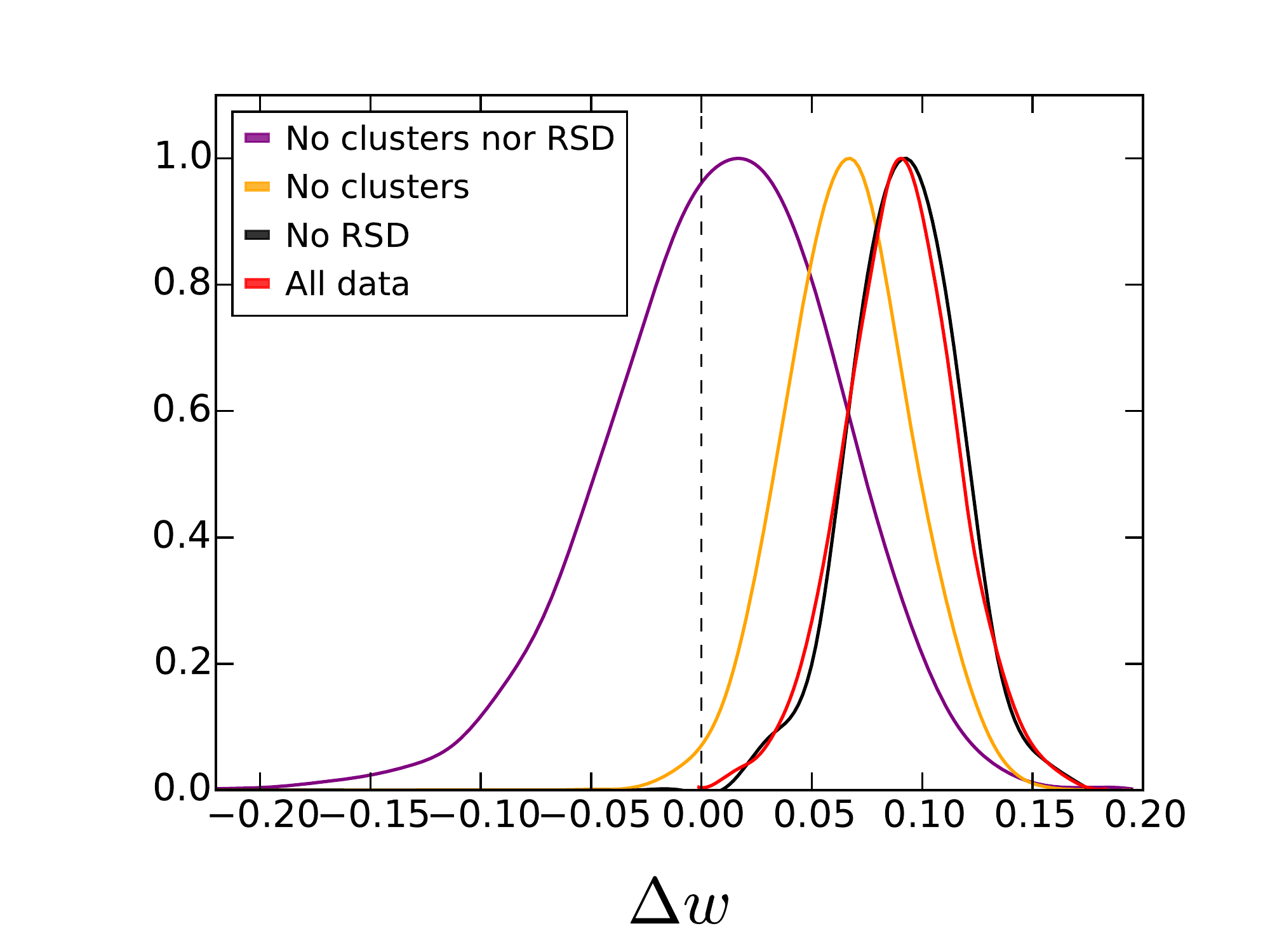}
\end{center}
\endminipage\hfill
\minipage{0.33\textwidth}
\begin{center}
\includegraphics[width=\textwidth]{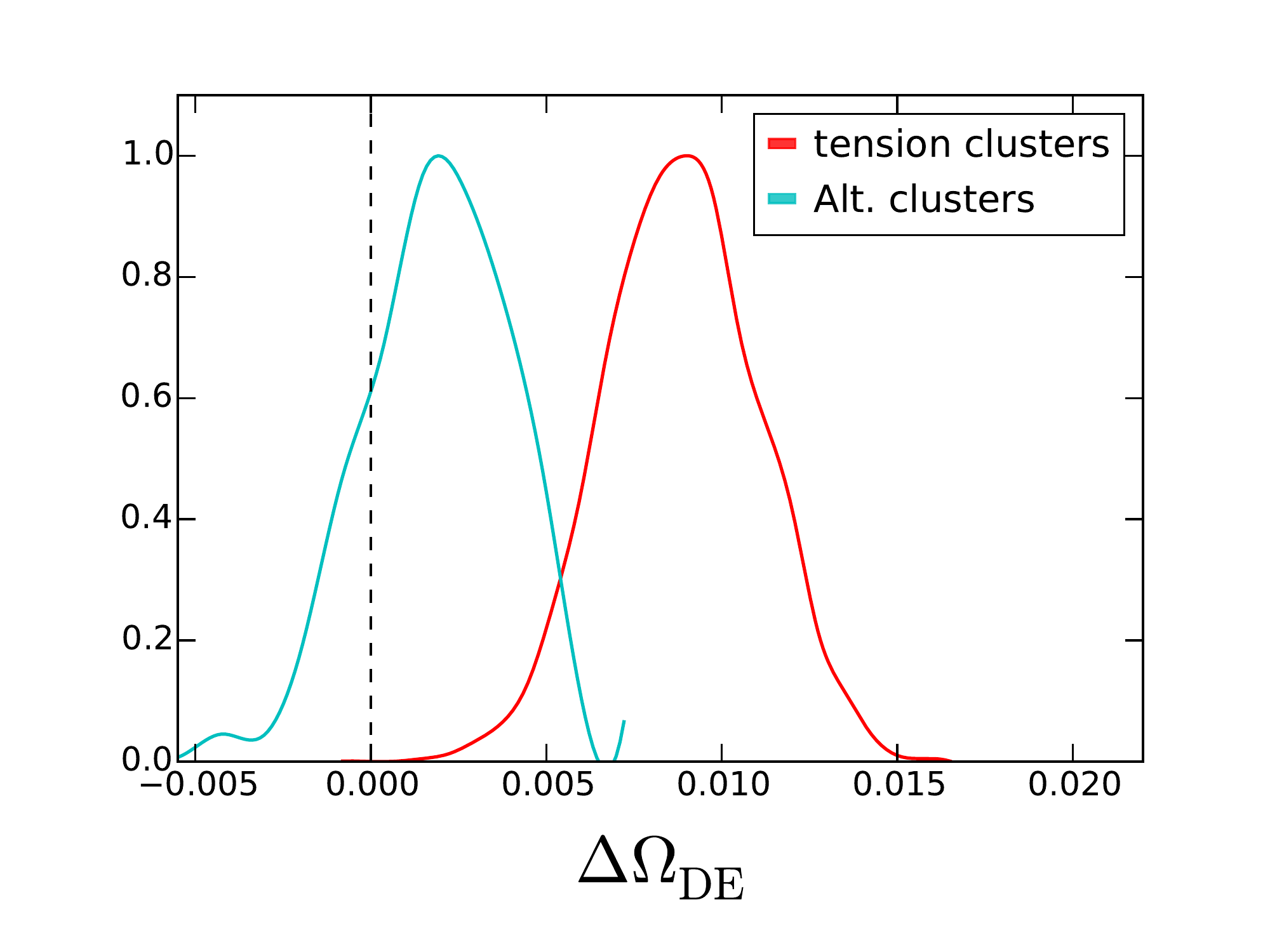}
\end{center}
\endminipage
\minipage{0.33\textwidth}
\begin{center}
\includegraphics[width=\textwidth]{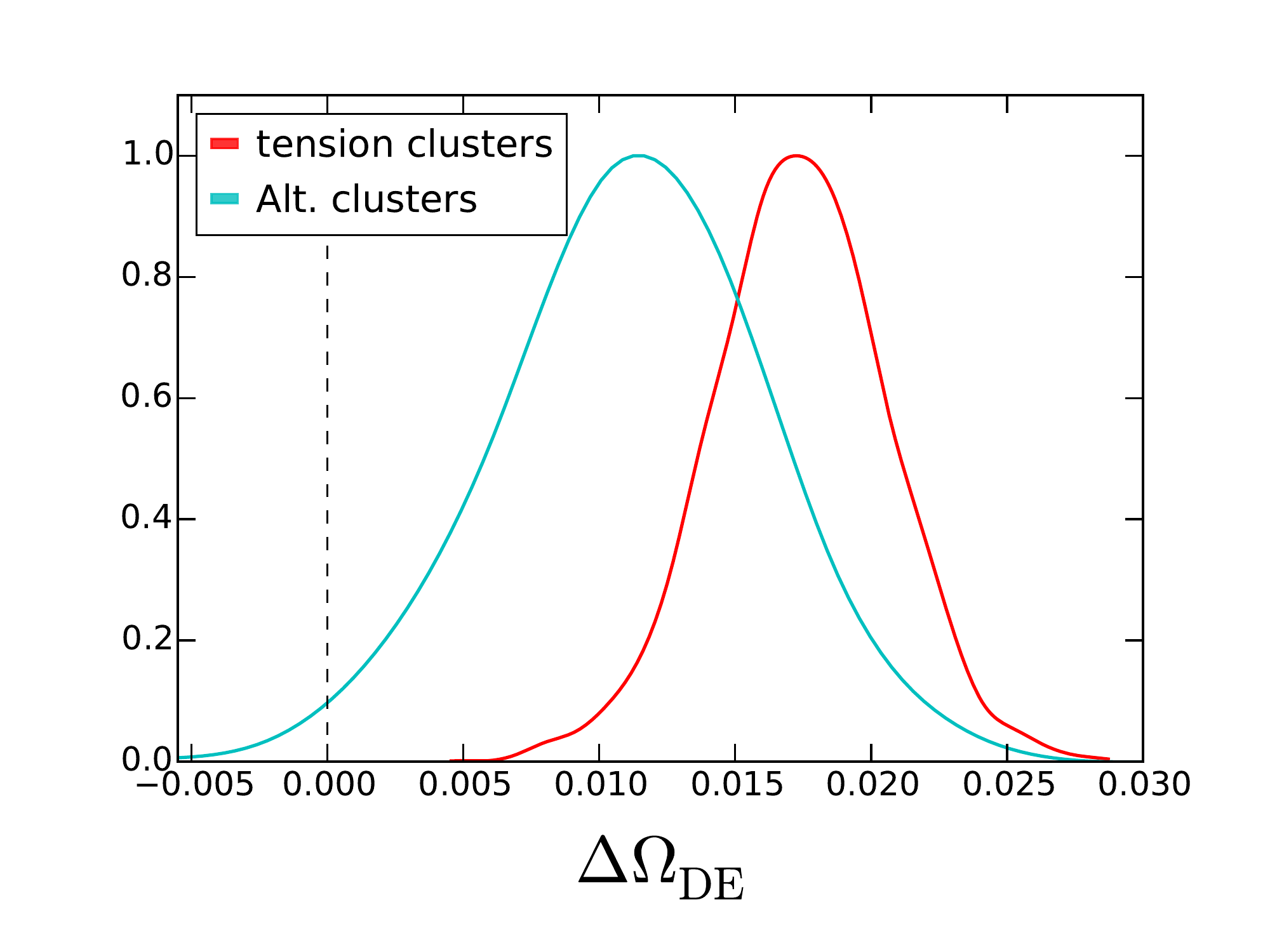}
\end{center}
\endminipage
\minipage{0.33\textwidth}
\begin{center}
\includegraphics[width=\textwidth]{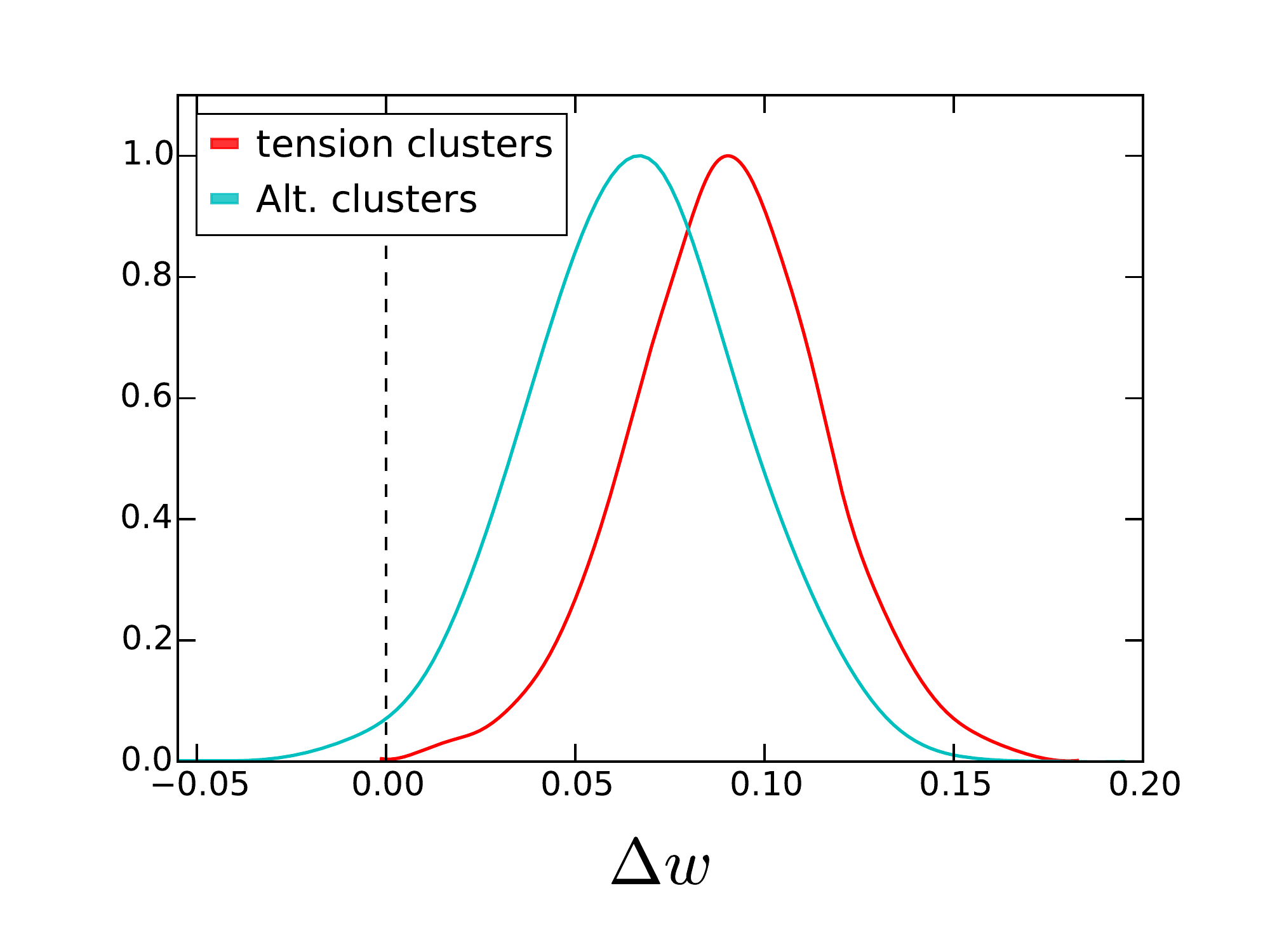}
\end{center}
\endminipage\hfill
\renewcommand{\baselinestretch}{1}
\caption{\footnotesize 
\textit{Top}: 68$\%$ and $95\%$  confidence level  for marginalized constraints in the  $\Omega_{\rm DE}^{\rm geom}$-$\Omega_{\rm DE}^{\rm growth}$ plane and $w_{\rm geom}$~-~$w_{\rm growth}$ plane.  Left panel:  splitting only   $\Omega_{\rm DE}$. Middle and right panels:  splitting both $w$ and $\Omega_{\rm DE}$. \textit{Center}:
 corresponding distributions for $\Delta \Omega_{\rm DE}$ and $\Delta w$. 
We show the constraints imposed by different combinations of data sets: using all of them (red, already shown in previous figures), removing RSD (black), removing cluster data set (orange) and removing both (purple). \textit{Bottom}: Marginalized distributions for $\Delta \Omega_{\rm DE}$ and $\Delta w$ using the all the data sets. The black dashed line indicates the points where the split parameters are equal. In red are shown the results using the clusters data set (already shown in previous figures) and in cyan, the alternative clusters data set.
}
\label{fig:datasets}
\end{figure}

The constraints imposed by the different data combinations are shown in the top and central panels of Figure \ref{fig:datasets}: tensions are mostly driven by the clusters data set, although for the cases where we split both $w$ and $\Omega_{\rm DE}$ (middle and right panels), RSD data  also contribute (the tension between the meta-parameters is $2.5\sigma$ in $w$ and $\Omega_{\rm DE}$ without including clusters data). However, using cluster abundance data and removing RSD data, the tension in the case of splitting only $\Omega_{\rm DE}$ is the same that the one obtained with all the data sets. Moreover, when we split both $w$ and $\Omega_{\rm DE}$ and make the analysis without including RSD data, the tensions are higher: 4.4$\sigma$ in $\Omega_{\rm DE}$ and 3.9 in $w$. If we change our clusters data set by the alternative clusters data set, the tensions are considerably reduced (in part due to the increased  error-bars): $1.3\sigma$ in $\Omega_{\rm DE}$ splitting only in this parameter and when we split in $\Omega_{\rm DE}$ and $w$, 2.6$\sigma$ and $2.5\sigma$, respectively (see bottom panels in Figure \ref{fig:datasets}).  When we combine our ``clusters data set"  and ``alternative clusters data set"  we find that, as expected, the  ``clusters data set" dominates and drives the fit.
In general clusters data require a lower amplitude of perturbations at late time and quasi-linear scales than that predicted by  the $\Lambda$CDM fit to CMB (and BAO) observations (see Figure \ref{fig:Preds_data}).  RSD data also follow this trend, but with reduced statistical significance. This can be appreciated in Figure \ref{fig:observables_growth} in the Appendix \ref{Appendix} (red and green lines in the two top panels). At present, there is not enough redundancy in the data to  clearly assess  whether this  effect is driven by new physics or by systematic effects. However, the significance of the deviation from the null hypothesis (i.e., the  difference in the meta-parameters) is large enough that if future investigations rule out a systematic effect  especially in clusters data, new physics in the GR/dark energy sector would have to be seriously considered.

%% file: sec6.tex
\section{Discussion and comparison with previous work}
\label{sec:discussion}

From the results of Sec.\ref{sec:splitbasemodel} it is 
 remarkable that we find significant differences when using Planck 2013 data instead of Planck 2015 data. The most surprising one is the disappearance of the tension in $w$ in the case of splitting both $w$ and $\Omega_{\rm DE}$
.
 For the problem at hand, the main difference between the  two CMB data sets is that $\tau_{\rm reio}$ is more tightly constrained by  Planck 2013  (due to the use of the low multipoles of polarization power spectrum of WMAP9) \citep{Planck15_likelihood}. As $\tau_{\rm reio}$ is less constrained  by Planck 2015, the CMB low multipoles constrain $w_{\rm growth}$ more  weakly and  cluster abundance data  (which need a lower $\sigma_8$ and thus  higher $w_{\rm growth}$) can therefore  drive $w_{\rm growth}$ away from $w_{\rm geom}=w_{\rm growth}$. 
On the other hand, WMAP9 low multipoles of polarization have more constraining power in $w_{\rm growth}$: the fit to the cluster data  therefore  drives  $\Omega_{\rm DE}^{\rm growth}$ higher.

\begin{figure}[h]
\minipage{0.5\textwidth}
\begin{center}
\includegraphics[width=0.9\textwidth]{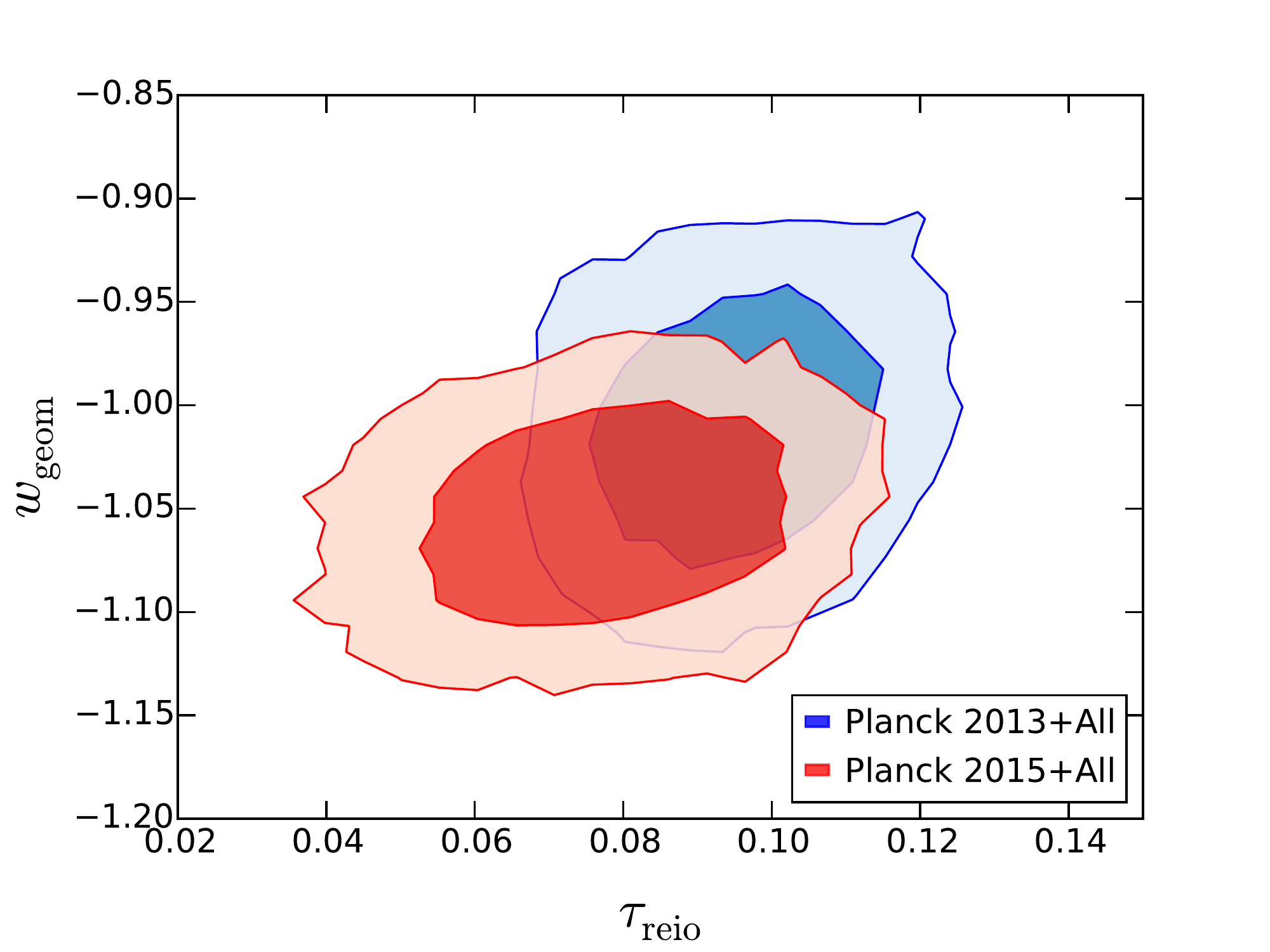}
\end{center}
\endminipage
\minipage{0.5\textwidth}
\begin{center}
\includegraphics[width=0.9\textwidth]{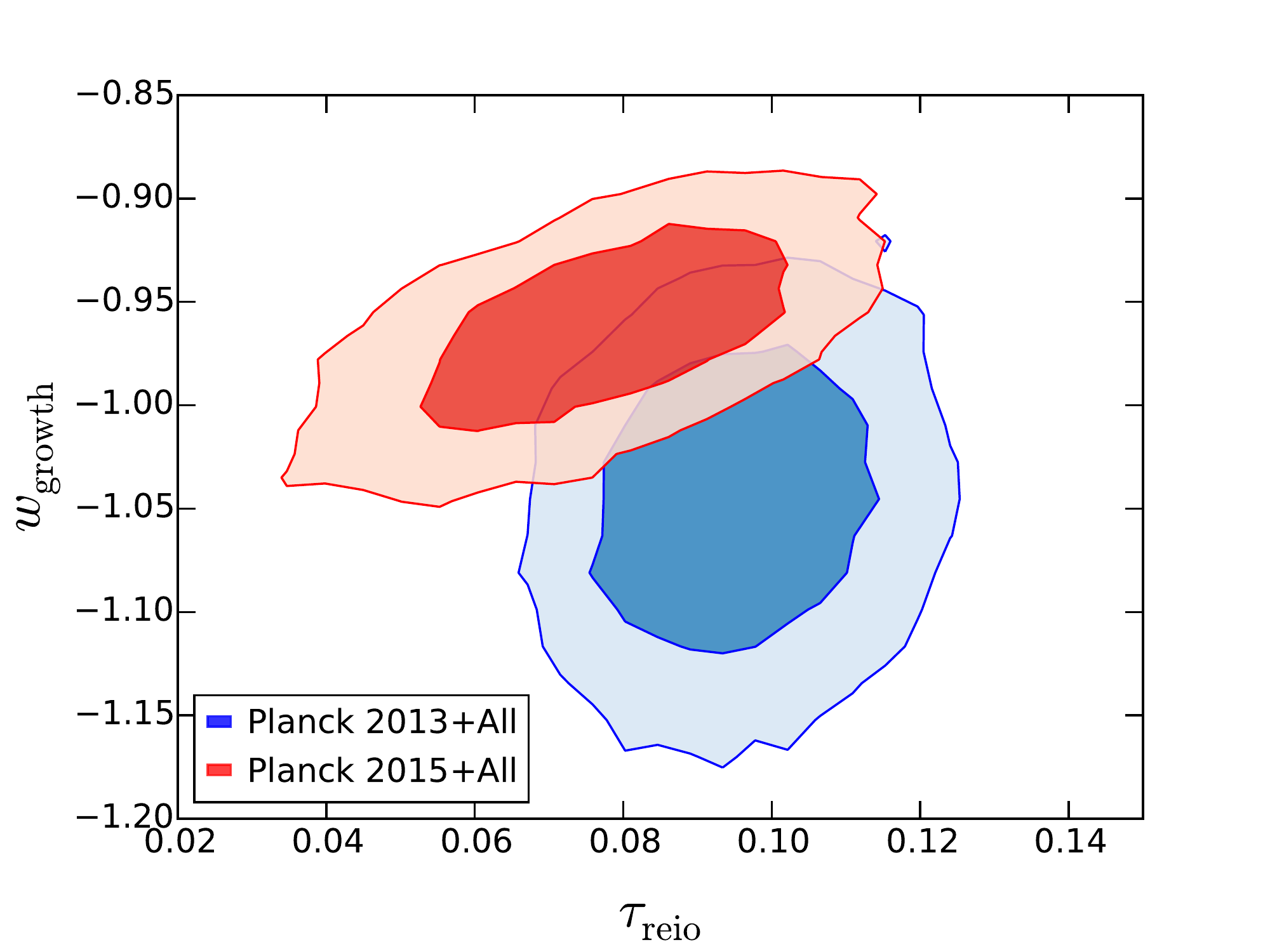}
\end{center}
\endminipage\hfill
\renewcommand{\baselinestretch}{1}
\caption{\footnotesize
68$\%$ and $95\%$ confidence level constraints in the $\tau_{\rm reio}$-$w$ plane for $w_{\rm geom}$ (\textit{left}) and $w_{\rm growth}$ (\textit{right}) in the case of splitting both $w$ and $\Omega_{\rm DE}$. With ``All" we refer to the data sets from all the cosmological probes we use but CMB (i.e. RSD, BAO, SNeIa and cluster abundance).
The results of the joint analysis using Planck 2013 data+All are shown in blue and the results using Planck 2015+All, in red. 
}
\label{fig:tau-w}
\end{figure}

 This  is illustrated in Figure \ref{fig:tau-w}, in the case of splitting both $w$ and $\Omega_{\rm DE}$. Using Planck 2015 data instead of Planck 2013 data has similar consequences in the rest of the cases. As noted elsewhere (e.g., \cite{Bellini15}), a  tighter constraint on $\tau_{\rm reio}$ through e.g., better E-mode polarization data at large scales, would  greatly help to reduce degeneracies and constrain  more tightly models beyond $\La$CDM.
 
There are also differences between our findings and  those of Ruiz et al. \citep{Ruiz15}. 
They use Planck 2013 and  find a tension in $w$ split parameters due to $w_{\rm growth}$, only when $w$ and $\Omega_{\rm DE}$ are split, mostly driven by RSD data.  We find a tension of a similar level in $w$ too, but only when we use Planck 2015 data. Moreover in our analysis, contrary to  Ref. \citep{Ruiz15}, it is the $\Omega_{\rm DE}^{\rm growth}$ meta-parameter that is affected most and it is driven by clusters data. This can be understood  as follows.

Beside the fact that the supernovae data sets are different in the two works, Ruiz et al. \citep{Ruiz15} use priors of the early universe (via the compressed constraints on the CMB peaks positions)  and do not include any  CMB constraint on growth from the low $\ell$ or lensing. We use the full Planck likelihood. For this reason our results are more similar to theirs  when we use the CMB data set that has the weakest constraints on $\tau_{\rm reio}$, as illustrated above.  For cluster abundance the volume factor in the halo mass function introduces a dependence on the geometry. Ref. \citep{Ruiz15} correctly accounts for this but they find that cluster abundance is only very weakly sensitive to geometry and mostly sensitive to growth\footnote{This statement  may ignore some higher dimensionality dependence of clusters on geometry. But this effect in any case would be small compared with the rest of the contributions considered in this work, both to geometry and growth.}. We therefore make the simplifying approximation that cluster abundance only constrains growth. 
For cluster data set Ref. \citep{Ruiz15} uses the   Rozo et al. \citep{Rozo10} data, which, as  it can be seen in Figure \ref{fig:Preds_data} (see also table \ref{tab:clusters}), is consistent with $\La$CDM and it is included in our ``alternative" clusters data set. We do not find any significant difference with respect to the results presented above if we use the same RSD data set  as  \citep{Ruiz15}. 

In the Appendix \ref{Appendix}  we show the observables -- $\Omega_{\rm M}^\beta \sigma_8$, $f\sigma_8$, temperature, polarization and lensing power spectra of CMB and the cross correlation of the  CMB temperature and polarization, BAO and SNeIa luminosity distance -- and the  corresponding theoretical predictions computed with the best fit parameters for each model.  Figure~\ref{fig:observables_geom}  shows  the probes of expansion history (where the theory prediction uses the geometry meta-parameters) and Figure \ref{fig:observables_growth} those  for the  growth of structures (the theory prediction uses the growth meta parameters). The color-coding goes as follows: $\La$CDM in blue, $w$ split in orange, $\Omega_{\rm DE}$ split in red and both split in green.  
BAO and SNeIa (top panels of Figure \ref{fig:observables_geom})  cannot discriminate among the different models (in the geometry sector) and it is the power spectrum of CMB (high multipoles, bottom panel of Figure~\ref{fig:observables_geom}) which  mostly constrains the geometry parameters. 
As already mentioned, cluster abundance is the main responsible for the tension between growth and geometry parameters. This is why the model in which  both parameters $w$ and $\Omega_{\rm DE}$ are split fits  the cluster data set  as well as the CMB data at all scales (green line and blue dots in the top left panel of Figure~\ref{fig:observables_growth}). Models with less freedom (i.e., less meta-parameters) do not fit so accurately. 

The $\La$CDM model   provides  a  better fit  to clusters and RSD than the model with $w$ only meta-parameters. However, to achieve this,  it does not fit as well the  high multipoles of temperature power spectrum of  the CMB.
The comparison between the predictions of the different models for the low multipoles of the cross correlation of temperature and polarization and the polarization power spectrum of CMB shows the effect of $\tau_{\rm reio}$.

%% file: sec7.tex
\section{Conclusions}\label{sec:conclusions}

We have performed an empirical  consistency test of GR/dark energy within a $w$CDM model by employing dark energy meta-parameters.  The test disentangles  expansion history and growth of structures constraints by replacing  each late-Universe parameter that describes the behavior of dark energy with two meta-parameters: one describing geometrical information in cosmological probes, and the other controlling the growth of structures. We refer to this procedure as parameter splitting. We have performed a  global analysis  using  state-of-the-art cosmological data sets of CMB,  SNeIa, BAO, cluster abundance and RSD.

Regardless of the parametrization chosen for geometry and growth, in the standard~cosmological model that assumes GR, the split parameters have to agree. This is the null hypothesis. Any disagreement  which excludes the null hypothesis at high statistical significance implies a violation of the standard cosmological model assumptions or unaccounted systematic errors.

Our work is  qualitatively consistent with previous similar works such as Ruiz et al \citep{Ruiz15},  keeping in mind the different methodology and the different data sets, and with previous claims of tensions between the value of $\sigma_8$ deduced from CMB data  and  the one measured in some  late time Universe observations. 

We find significant tensions between the split parameters whenever we split $\Omega_{\rm DE}$ (3.8$\sigma$ splitting only $\Omega_{\rm DE}$  and   3.5$\sigma$  in $w$  and $>4.4\sigma$ in $\Omega_{\rm DE}$ when  splitting both parameters). These tensions are only partially alleviated  when  the sum of neutrino masses is allowed to be a free parameter, but in this case the required value of $\sum m_\nu$ is uncomfortably high given  the  current  cosmological upper limits.

Improved S/N for CMB polarization low multipoles would reduce the  residual degeneracy between $\tau_{\rm reio}$ and $w_{\rm growth}$ which appears to be driving  $w_{\rm growth}>w_{\rm geom}$ in some cases. On the other hand  the tension in $\Omega_{\rm DE}$ is more significant and persistent. 

We identify a specific  set of measurements of cluster abundances, those obtained from X-ray observations of galaxy clusters made by Chandra \citep{Vikhlinin09} and from the Sunyaev-Zeldovich effect measured by Planck \citep{Planck13_SZ},  as the main responsible of the tensions. 
Using instead an alternative set composed by 
six measurements of cluster abundance (see table \ref{tab:clusters}),  we find no significant violation of the null hypothesis; however  the statistical power of this data set is  lower and  the error-bars are larger. At present there is not enough redundancy in the data  to enable us to analyze separately different sub-sets of data (i.e., clusters data) to investigate possible systematics.  
The significance of the difference between meta-parameters, however, is high enough that, should systematic effects be ruled out,   new physics in the GR/dark energy sector would have to be seriously considered.

 This leads us to conclude that, before interpreting the tension as a failure of the GR+$w$CDM  model, a better modeling and interpretation of cluster abundance as a probe of the growth of cosmic structures is needed.

We envision that consistency tests such as the one presented here will be valuable for forthcoming cosmological data given their  increased (forecasted) precision.

%% file: Acknowledgements.tex
We thank Dragan Huterer for valuable comments on an early version of the draft which help improve the paper.
JLB is supported by the Spanish MINECO under grant BES-2015-071307, co-funded by the ESF. JLB acknowledges financial support from CEI UAM-CSIC during part of this work.
LV and AJC are supported by the European Research Council under the European Community's Seventh Framework Programme FP7-IDEAS-Phys.LSS 240117. Funding for this work was partially provided by the Spanish MINECO under projects AYA2014-58747-P and MDM-2014-0369 of ICCUB (Unidad de Excelencia ``Maria de Maeztu"). LV acknowledges hospitality of Imperial  Center for Inference and Cosmology   and  support by Royal Society grant IE140357. AJC acknowledges hospitality of ITC, Harvard-Smithsonian Center for Astrophysics, Harvard University.

Based on observations obtained with Planck (http://www.esa.int/Planck), an ESA science mission with instruments and contributions directly funded by ESA Member States, NASA, and Canada.

Funding for SDSS-III has been provided by the Alfred P. Sloan Foundation, the Participating Institutions, the National Science Foundation, and the U.S. Department of Energy Office of Science. The SDSS-III web site is http://www.sdss3.org/.

SDSS-III is managed by the Astrophysical Research Consortium for the Participating Institutions of the SDSS-III Collaboration including the University of Arizona, the Brazilian Participation Group, Brookhaven National Laboratory, Carnegie Mellon University, University of Florida, the French Participation Group, the German Participation Group, Harvard University, the Instituto de Astrofisica de Canarias, the Michigan State/Notre Dame/JINA Participation Group, Johns Hopkins University, Lawrence Berkeley National Laboratory, Max Planck Institute for Astrophysics, Max Planck Institute for Extraterrestrial Physics, New Mexico State University, New York University, Ohio State University, Pennsylvania State University, University of Portsmouth, Princeton University, the Spanish Participation Group, University of Tokyo, University of Utah, Vanderbilt University, University of Virginia, University of Washington, and Yale University.

%% file: Appendix.tex
\appendix
\section{Appendix}\label{Appendix}

\begin{figure}[h]
\begin{center}
\includegraphics[width=0.6\textwidth]{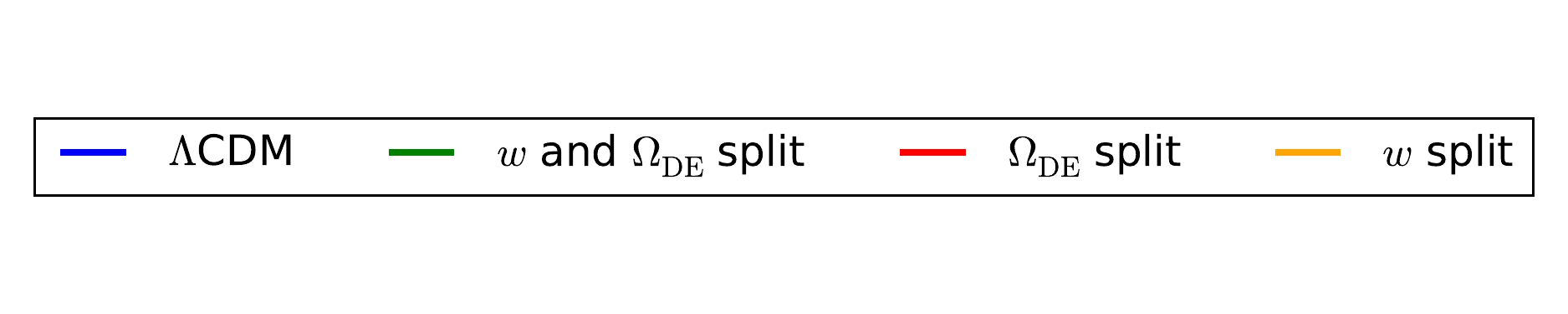}
\end{center}
\hfill
\minipage{0.5\textwidth}
\begin{center}
\includegraphics[width=0.9\textwidth]{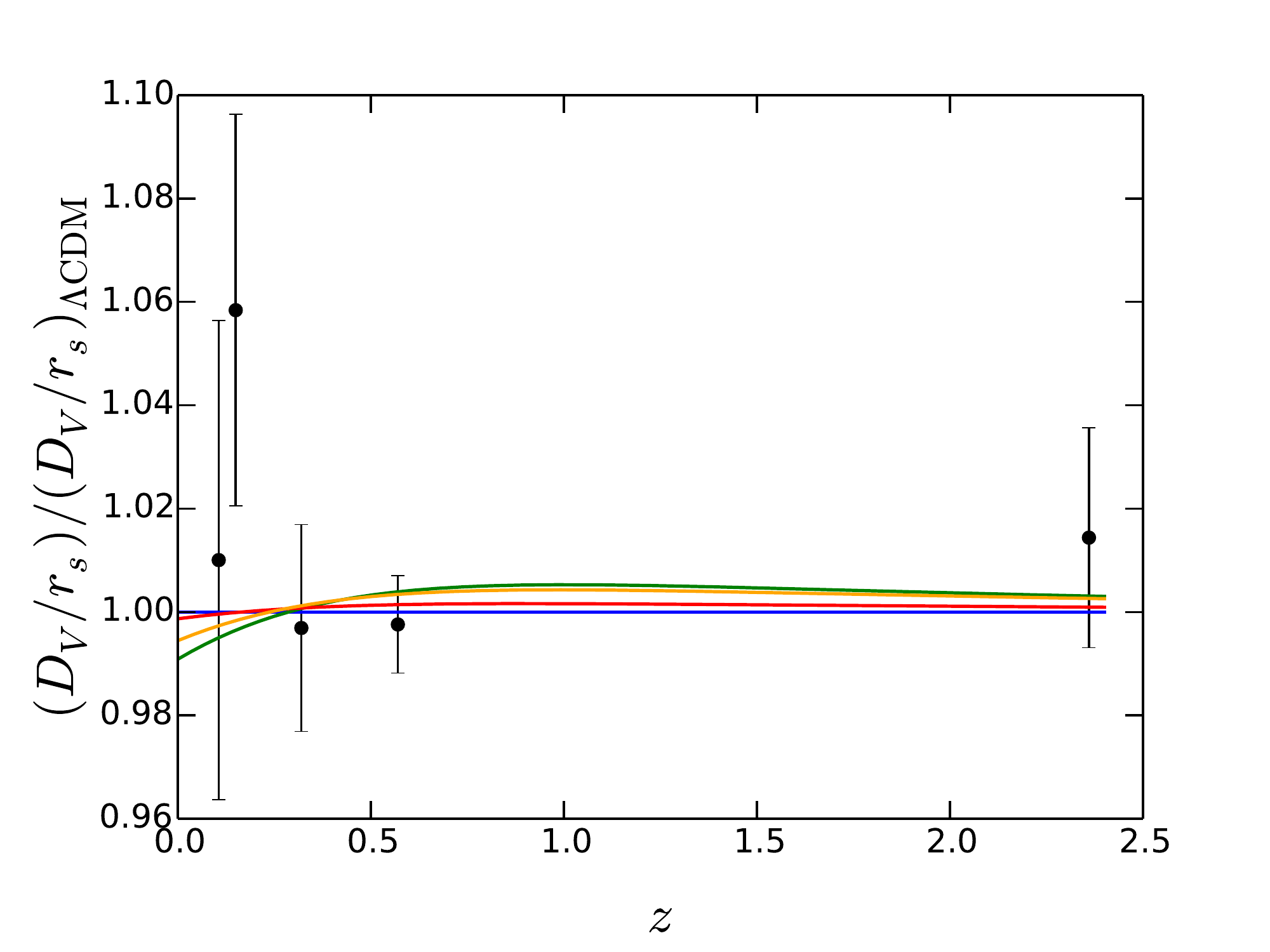}
\end{center}
\endminipage
\minipage{0.5\textwidth}
\begin{center}
\includegraphics[width=0.9\textwidth]{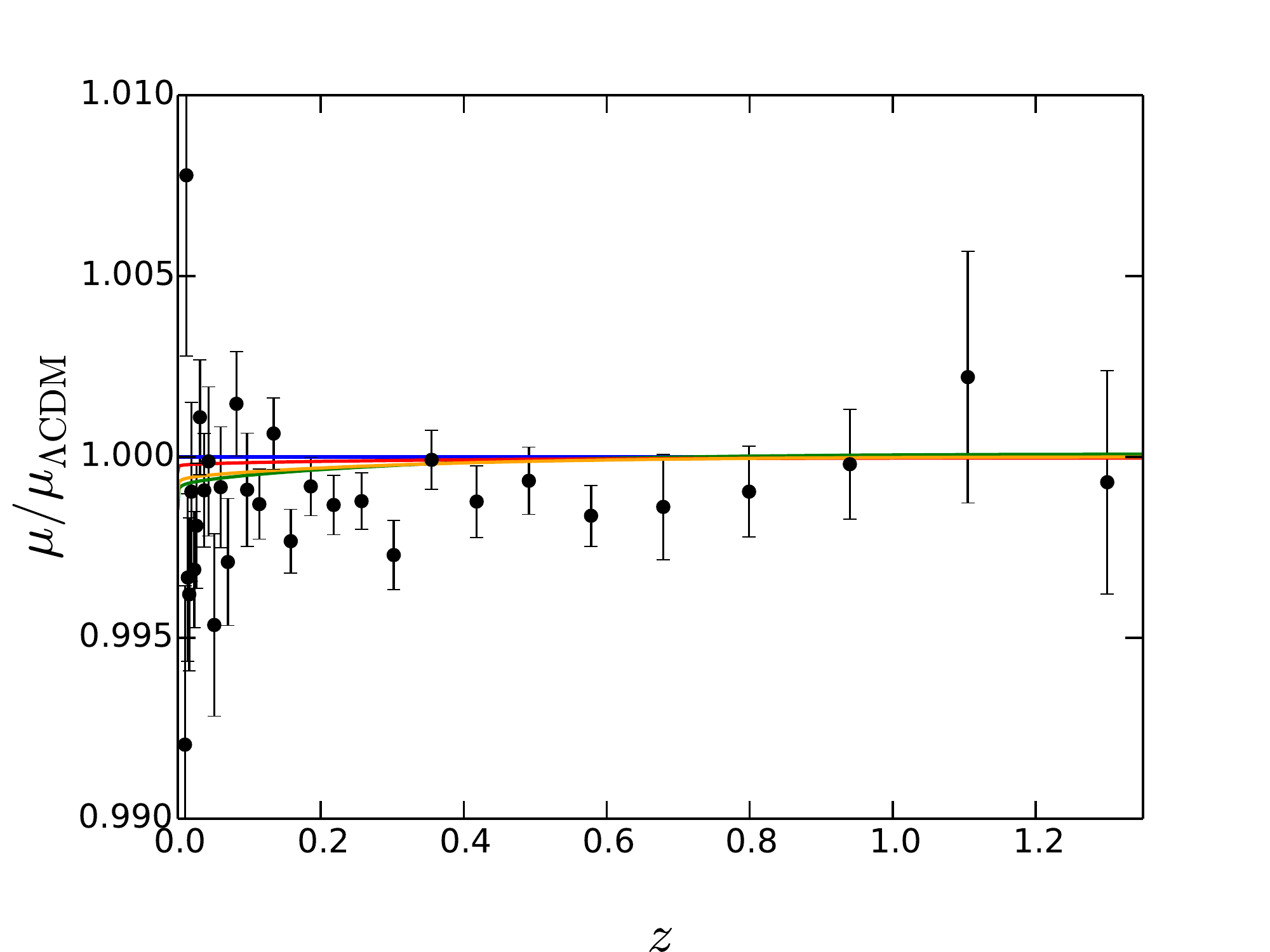}
\end{center}
\endminipage\hfill
\begin{center}
\minipage{0.5\textwidth}
\begin{center}
\includegraphics[width=0.9\textwidth]{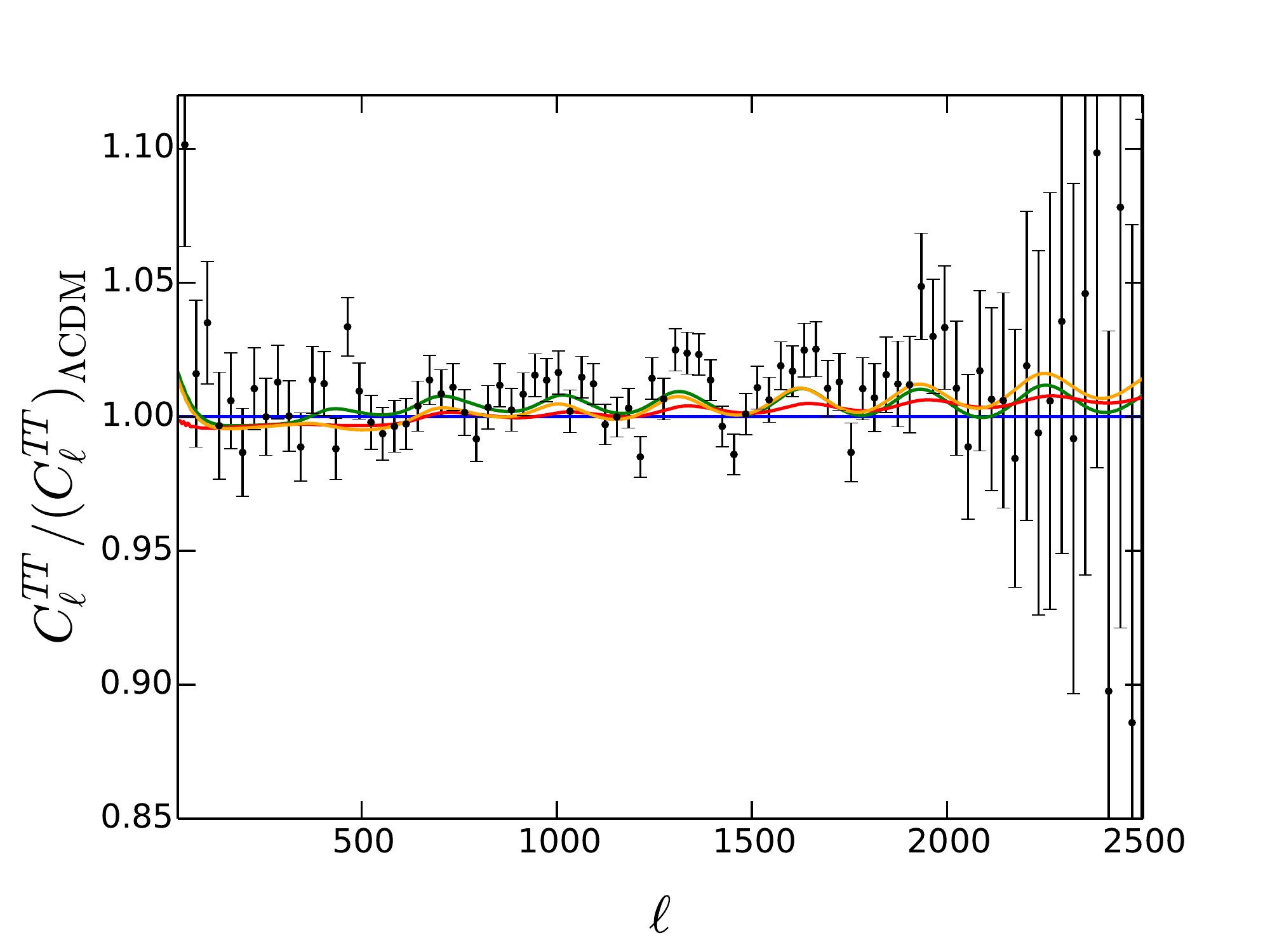}
\end{center}
\endminipage\hfill
\end{center}
\renewcommand{\baselinestretch}{1}
\caption{\footnotesize
Data sets used in this work to constrain the background together with the best fit  theoretical predictions  in the following cases:
 $w$ split (orange), $\Omega_{\rm DE}$ split (red),  two-parameters split (green) and $\La$CDM (blue). Theoretical predictions are computed with geometry meta-parameters. The data are: $D_V/r_s$ from BAO (top left), $\mu =5\log_{10}(D_L/10$pc$)$
 from SNeIa (top right) and the high multipoles of the CMB  temperature power spectrum (bottom). The point from the Ly$\alpha$ measurements of BAO ($z=2.36$, top left panel) is computed from the observed $H r_s$ and $D_A/r_s$ as $(\frac{c}{H})^{0.7}D_A^{0.3}/r_s$, following \citep{Delubac15}.  Quantities are normalized by the prediction of a $\La$CDM model with the best fit  parameters obtained from our analysis.}

\label{fig:observables_geom}
\end{figure}

\begin{figure}[h]
\begin{center}
\includegraphics[width=0.6\textwidth]{Figures/legend.pdf}
\end{center}
\hfill
\minipage{0.5\textwidth}
\begin{center}
\includegraphics[width=0.9\textwidth]{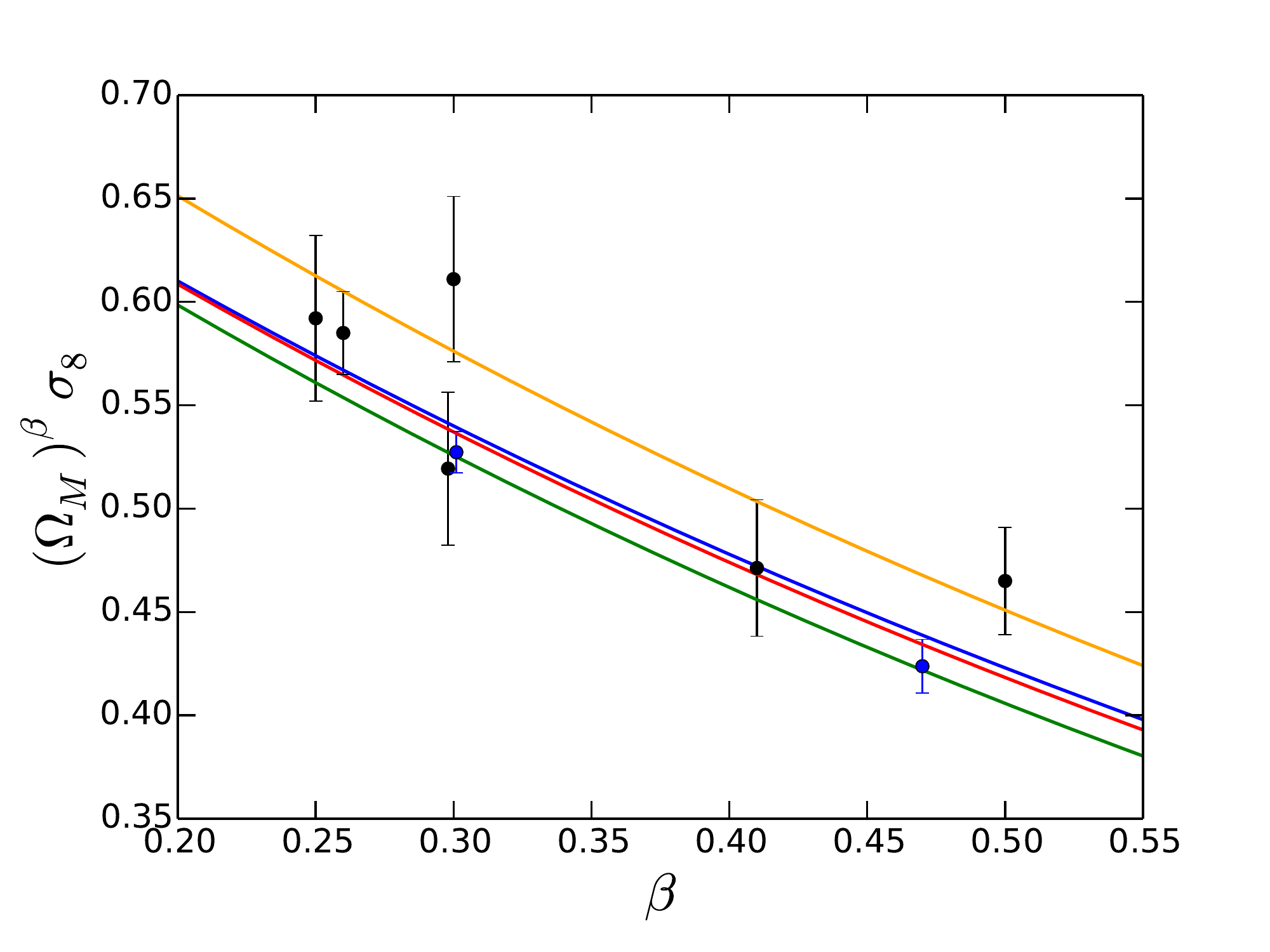}
\end{center}
\endminipage
\minipage{0.5\textwidth}
\begin{center}
\includegraphics[width=0.9\textwidth]{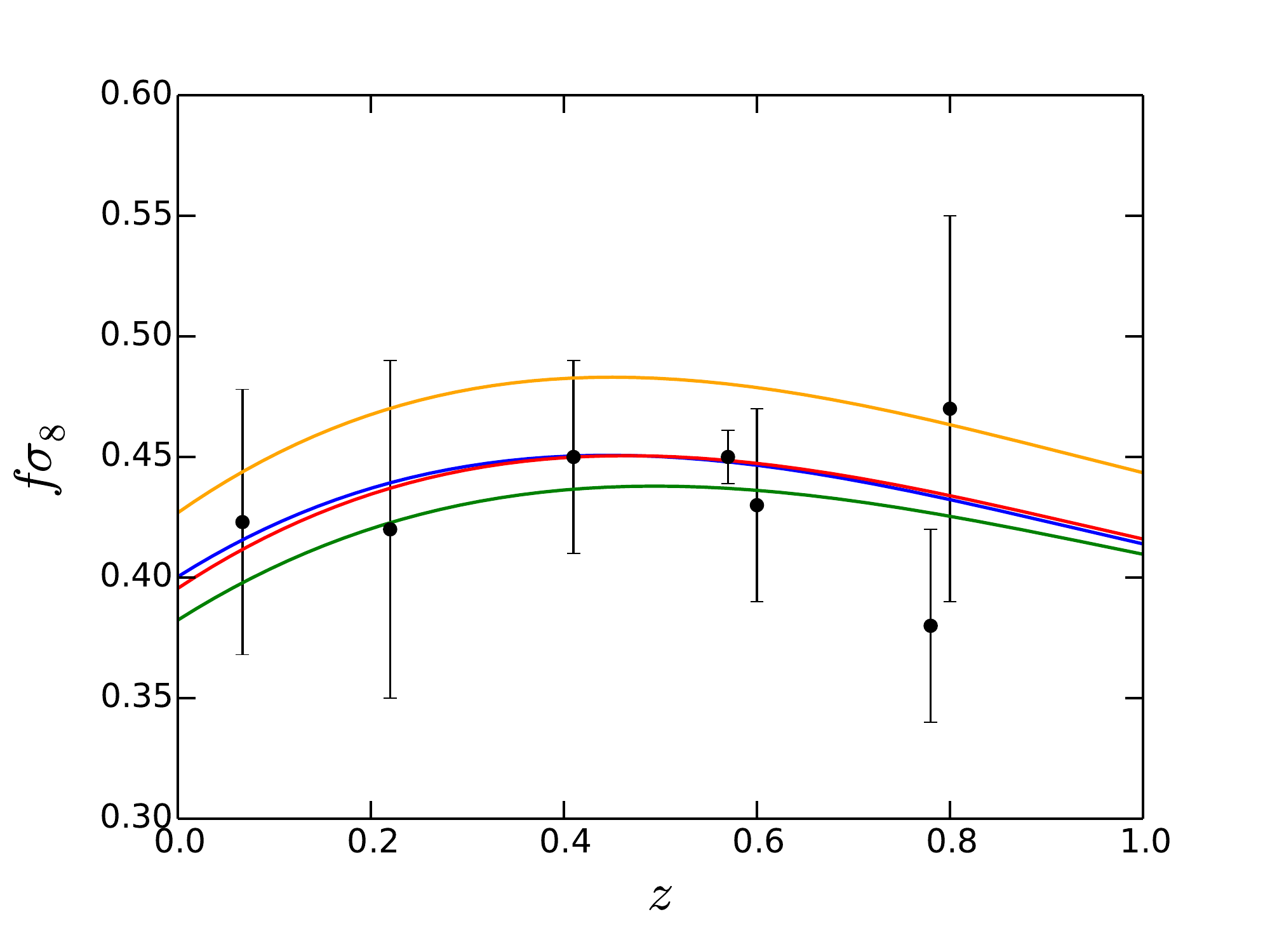}
\end{center}
\endminipage\hfill
\minipage{0.5\textwidth}
\begin{center}
\includegraphics[width=0.9\textwidth]{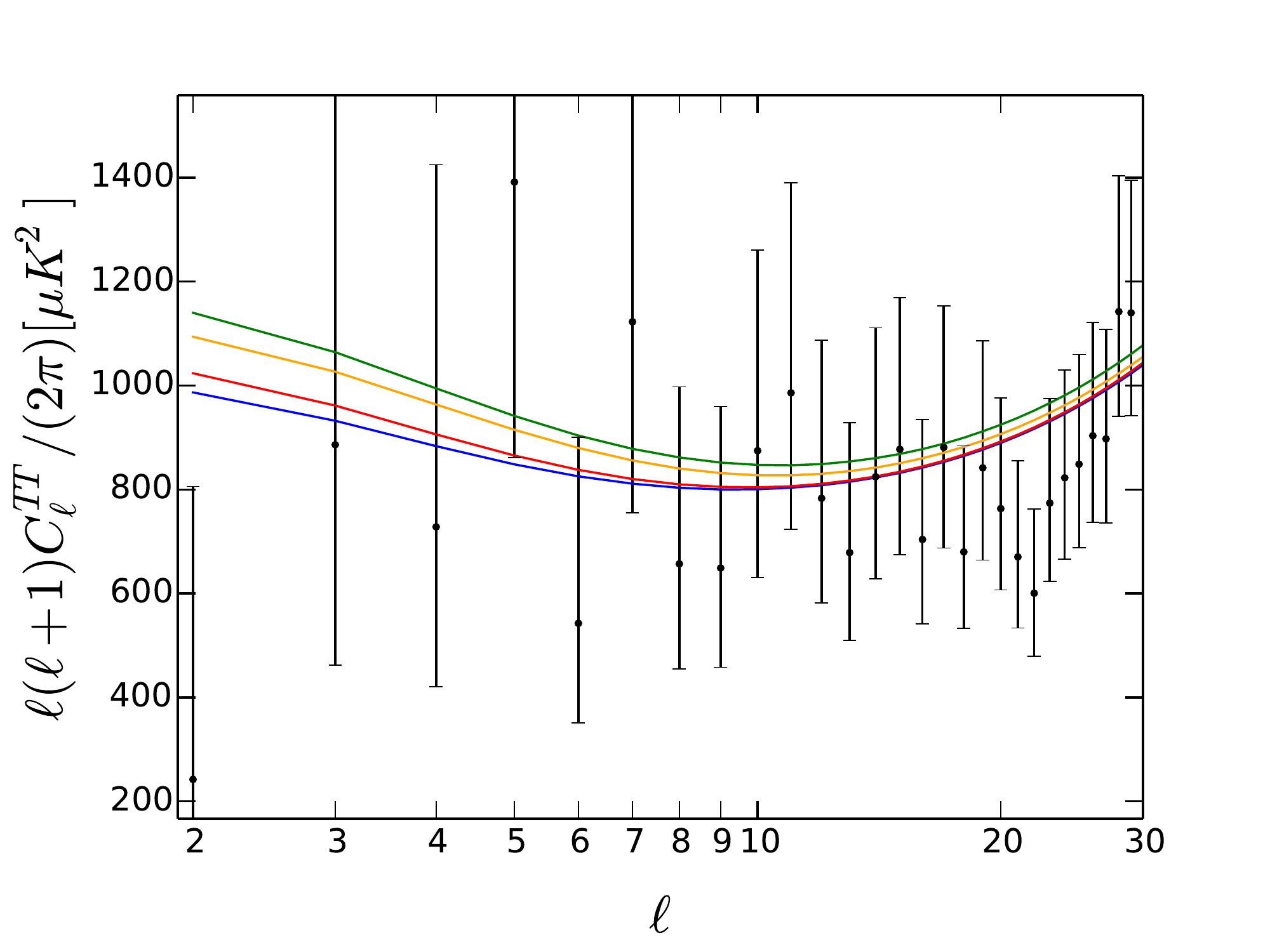}
\end{center}
\endminipage
\minipage{0.5\textwidth}
\begin{center}
\includegraphics[width=0.9\textwidth]{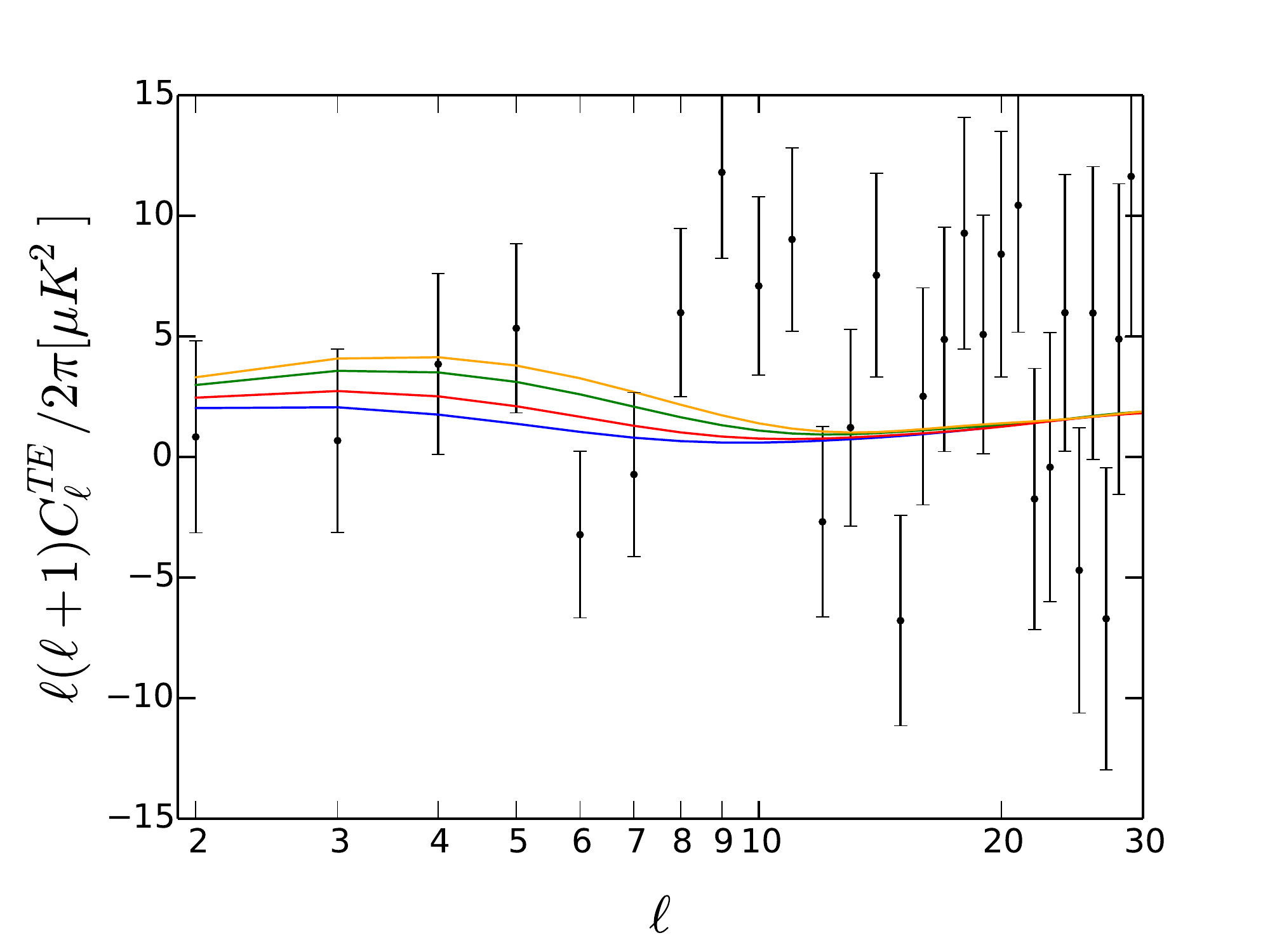}
\end{center}
\endminipage\hfill
\minipage{0.5\textwidth}
\begin{center}
\includegraphics[width=0.9\textwidth]{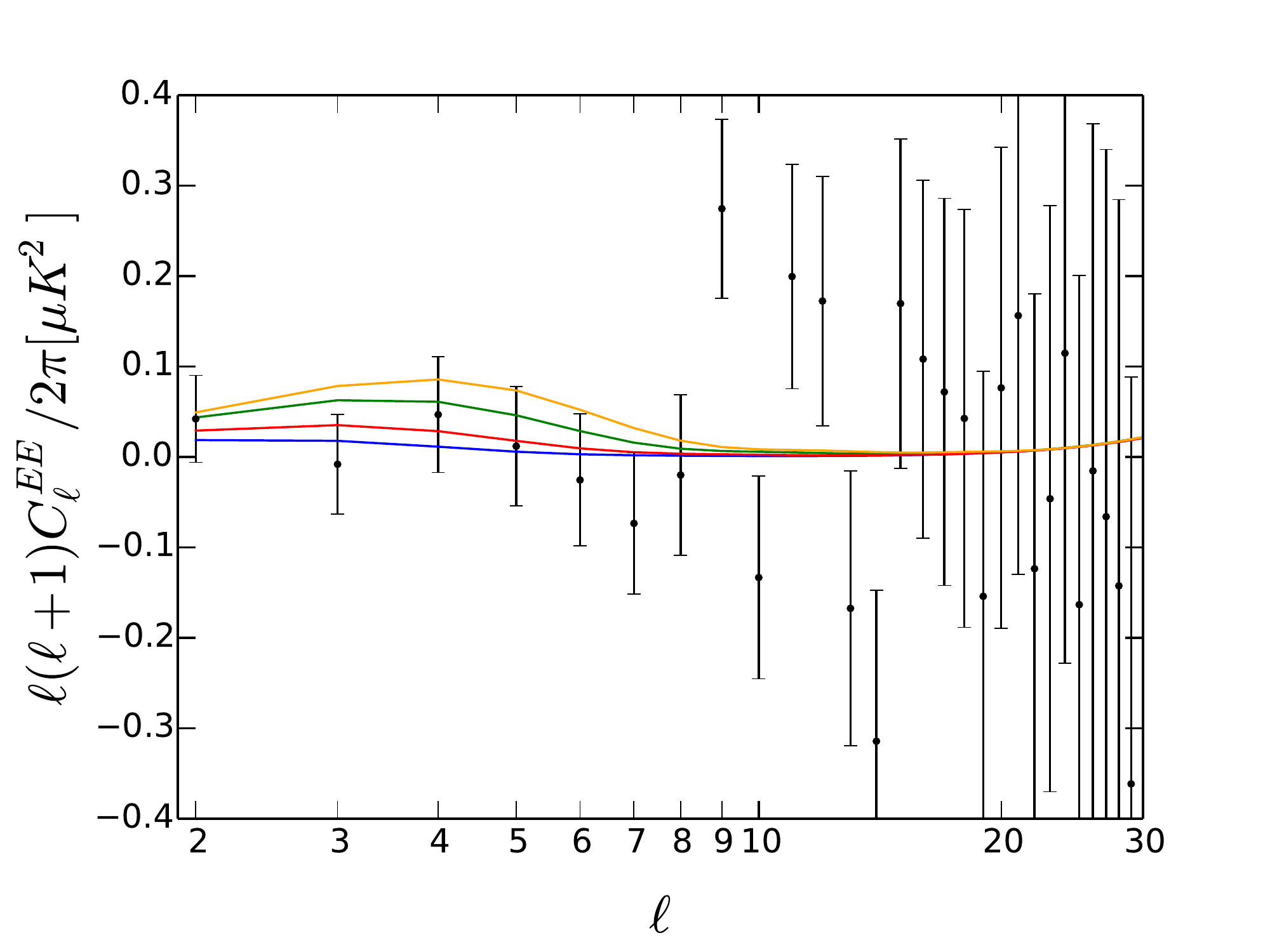}
\end{center}
\endminipage
\minipage{0.5\textwidth}
\begin{center}
\includegraphics[width=0.9\textwidth]{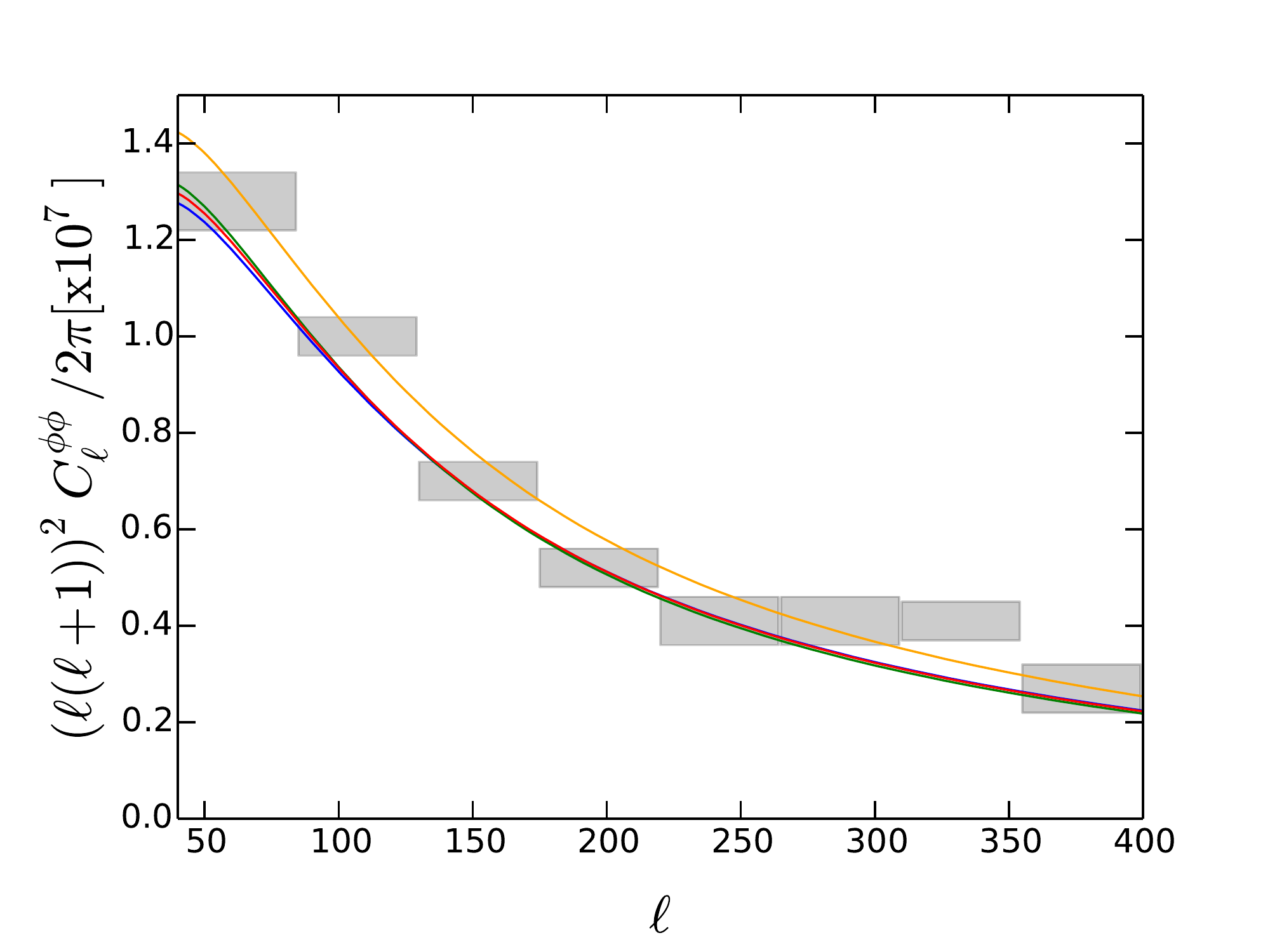}
\end{center}
\endminipage\hfill
\renewcommand{\baselinestretch}{1}
\caption{\footnotesize
Data sets used  to constrain the growth history together with the best fit  theoretical predictions  in the following cases:
 $w$ split (orange), $\Omega_{\rm DE}$ split (red),  two-parameters   split (green) and $\La$CDM (blue). Theoretical predictions are computed with growth meta-parameters. The data are: $\Omega_{\rm M}^\beta\sigma_8$ from cluster abundance (top left), $f\sigma_8$ from RSD (top right) and the low multipoles of the temperature power spectrum (central left),the cross correlation of temperature and polarization (central right), the polarization power spectrum (bottom left) and the lensing potential power spectrum (bottom right) of CMB. 
}
\label{fig:observables_growth}
\end{figure}

It is useful for the interpretation of our results to compare the  data with the best fit  theory line  for different cases: no meta-parameters,  only one meta-parameter or splitting  both  dark energy parameters. This is what we report here and illustrate in Figures \ref{fig:observables_geom} and \ref{fig:observables_growth}.  In some cases (BAO, SNe, high-$\ell$ CMB temperature power spectrum) we normalize all quantities to the best fit $\La$CDM model for clarity.

Clearly clusters data and, to lower statistical significance, RSD data, require a lower amplitude of fluctuations than predicted by a standard $\La$CDM (e.g.,  with parameters determined by  a fit to CMB-only or CMB+BAO data)  or $w$CDM model.

